\documentclass[acmsmall]{acmart}

\usepackage[english]{babel}
\usepackage{amsmath}
\usepackage{cancel}
\usepackage{xspace}
\usepackage{xcolor}
\usepackage{enumitem}  
\usepackage{pifont}   
\usepackage{algorithm}
\usepackage{algpseudocode}
\usepackage[acronym]{glossaries}
\usepackage{booktabs}
\usepackage{tabularx}
\usepackage{multirow}
\usepackage{makecell} 
\usepackage{wrapfig}
\usepackage[labelfont=bf]{caption}
\usepackage{subcaption}
\usepackage[T1]{fontenc} 
\usepackage{url}
\usepackage{hyperref}

\newcommand{\cmark}{\ding{51}}   
\newcommand{\xmark}{\ding{55}}   
\newcommand{\pmark}{\raisebox{0.2ex}{\boldmath$\sim$}}  

\newcommand{\slicescope}{\emph{SliceScope}\xspace}
\newcommand{\E}{\mathbb{E}}

\newcommand{\dint}{\emph{Static slice-agnostic}\xspace}
\newcommand{\dintst}{\emph{Static slice-aware}\xspace}

\newcommand{\eat}[1]{}
\newcommand{\signpost}[1]{\noindent \textbf{#1.}}
\newcommand{\indsignpost}[1]{\textbf{#1.}}

\newcommand{\slaval}{\alpha}
\newcommand{\knob}{k_{s,m}}
\newcommand{\dknob}{\Delta_{s,m}}
\newcommand{\cknob}{\delta_{s,m}}

\newcolumntype{Y}{>{\centering\arraybackslash}X} 
\newcolumntype{A}{>{\centering\arraybackslash}p{2cm}}
\newcolumntype{B}{>{\centering\arraybackslash}p{1.4cm}}
\newcolumntype{C}{>{\centering\arraybackslash}p{2.0cm}}
\newcolumntype{D}{>{\centering\arraybackslash}p{2.3cm}} 

\newacronym{SLO}{SLO}{Service Level Objective}
\newacronym{SLA}{SLA}{Service Level Agreement}
\newacronym{VPN}{VPN}{Virtual Private Network}
\newacronym{QoS}{QoS}{Quality of Service}
\newacronym{INT}{INT}{In-band Network Telemetry}
\newacronym{URLLC}{URLLC}{Ultra Reliable Low Latency Communications}
\newacronym{eMBB}{eMBB}{Enhanced Mobile Broadband}
\newacronym{mMTC}{mMTC}{Massive Machine Type Communications}
\newacronym{NF}{NF}{Network Function}
\newacronym{SDP}{SDP}{Service Demarcation Point}
\newacronym{3GPP}{3GPP}{3rd Generation Partnership Project}
\newacronym{RAN}{RAN}{Radio Access Network}
\newacronym{KPI}{KPI}{Key Performance Indicator}
\newacronym{KPM}{KPM}{Key Performance Metric}
\newacronym{E2E}{e2e}{end-to-end}
\newacronym{E2H}{E2H}{End-to-Hop}
\newacronym{IETF}{IETF}{Internet Engineering Task Force}
\newacronym{GTP}{GTP}{GPRS Tunnelling Protocol}
\newacronym{TEID}{TEID}{Tunnel Endpoint Identifier}
\newacronym{SRv6}{SRv6}{Segment Routing IPv6}
\newacronym{ILP}{ILP}{Integer Linear Program}
\newacronym{PSTO}{PSTO}{Per-slice Telemetry Optimization}
\newacronym{MNO}{MNO}{Mobile Network Operator}
\newacronym{MPLS}{MPLS}{Multiprotocol Label Switching}

\setcopyright{none}   

\acmVolume{}
\acmNumber{}
\acmArticle{}

\renewcommand\footnotetextcopyrightpermission[1]{}

\newif\ifanon

\title[SliceScope]{Dynamic SLA-aware Network Slice Monitoring}

\ifanon
  \author{Submission \#76}
\else
  \author{Niloy Saha}
  \affiliation{%
    \institution{University of Waterloo}
    \country{Canada}
  }
  \email{n6saha@uwaterloo.ca}
  \author{Mina Tahmasbi Arashloo}
  \affiliation{%
    \institution{University of Waterloo}
    \country{Canada}
  }
  \email{mina.arashloo@uwaterloo.ca}
  \author{Nashid Shahriar}
  \affiliation{%
    \institution{University of Regina}
    \country{Canada}
  }
  \email{nashid.shahriar@uregina.ca}
  \author{Raouf Boutaba}
  \affiliation{%
    \institution{University of Waterloo}
    \country{Canada}
  }
  \email{rboutaba@uwaterloo.ca}
  \renewcommand{\shortauthors}{Saha et al.}
\fi

\begin{document}

\begin{abstract}
Next-generation networks increasingly rely on \textit{network slices} \textemdash{} logical networks tailored to specific application requirements, each with distinct Service-Level Agreements (SLAs). Ensuring compliance with these SLAs requires continuous, real-time monitoring of end-to-end performance metrics for each slice, within a limited telemetry budget. However, we find that existing solutions face two fundamental limitations: they either lack end-to-end visibility (e.g., sketches, probabilistic sampling) or provide visibility but lack the control mechanisms to dynamically allocate monitoring resources according to slice SLAs.

We address this through a formal framework that reframes slice monitoring as a \emph{closed-loop control problem}, and defines the minimal data plane requirements for SLA-aware slice monitoring via a \textit{telemetry primitive contract}. We then present \slicescope, a realization of this framework that combines:
(1) a control strategy that dynamically allocates the monitoring resources across diverse slices according to their SLA criticality, and (2) a data-plane based on change-triggered INT that provides per-packet end-to-end visibility with tunable accuracy-overhead trade-offs, satisfying the telemetry contract.
Our evaluation results on programmable switches and in large-scale simulations with a mixture of different slice types, demonstrate that \textit{\slicescope tracks critical slices up to $4\times$ more accurately} compared to static baselines, while showing that \textit{change-triggered INT outperforms alternative primitives} for realizing the telemetry primitive contract.

\end{abstract}

\maketitle

\section{Introduction} \label{sec:introduction}


A network slice is an isolated logical network that provides specific capabilities and performance characteristics to an application or service over a shared physical network. These guarantees are typically expressed through \glspl{SLA} over metrics such as throughput, latency, jitter, and loss. 
Modern networks can have dozens to hundreds of concurrently active slices \cite{atis-slicing, balasingam24, mittal-slicing}, each with distinct \glspl{SLA}.
For example, in the context of 5G networks, Ultra Reliable Low Latency Communication (URLLC) slices are expected to provide very low loss rates (e.g., $<0.001\%$) and end-to-end latency (e.g., 1--5 ms) but have relatively flexible bandwidth requirements (e.g., 1--10 Mbps)~\cite{3gpp_22.261, 5gamericas2017, 5gamericas2021}. Enhanced Mobile BroadBand (eMBB) slices, by contrast, have higher bandwidth requirements (e.g., up to $1000$ Mbps) but tolerate higher packet loss (e.g., $<1\%$) and end-to-end latency (e.g., 10--50 ms) \cite{3gpp_22.261, 5gamericas2017, 5gamericas2021}. 

To ensure compliance with these SLAs, network operators need to \emph{continuously} monitor, in \emph{real time}, the relevant SLA metrics for \emph{every active slice}.
This is essential for timely detection of SLA violations -- operators must know immediately if the per-packet latency of a latency-critical slice exceeds the acceptable threshold to trigger appropriate mitigation actions.
%
To see why this is challenging, 
consider a 5G network that can support 500 active slices. While slices can be broadly categorized into the URLLC and eMBB types, each instance comes with its own unique SLA that specifies its concrete requirements in terms of end-to-end latency, loss rate, and throughput. Moreover, the set of active slices \emph{dynamically changes over time} -- service providers using the 5G network as their substrate can request new slices to be created or existing ones to be removed as needed.
The operators of the 5G network need to monitor each slice to ensure that its traffic, in aggregate, meets the unique performance requirements specified in that slice's SLA.

%
For instance, if 10 out of the 500 slices have very strict SLAs for end-to-end (e2e) latency, the operator could decide, for example, to use a packet-level telemetry mechanism such as In-band Network Telemetry (INT)~\cite{int}, which embeds measurement data within packet headers, to monitor end-to-end latency. They could then collect INT information on every packet belonging to those 10 slices but less frequently for the rest, keeping the total overhead within budget.

Now, suppose the set of active slices changes such that 100 out of the 500 slices have very strict SLAs for e2e latency, and collecting telemetry information for every packet of all these 100 slices would exceed the total monitoring budget. In this case, 
the best strategy to allocate this limited budget could be to collect telemetry information from critical slices at a ``base'' frequency, but adjust it per slice if e2e latency fluctuations in a particular slice require more frequent INT collection.
That is, in the presence of a dynamically changing set of slices, there is no ``one-size-fits-all'' monitoring strategy -- as the set of active slices, their SLAs, and their traffic patterns change, the monitoring system needs to continuously revisit how it allocates its monitoring budget among the active slices in the network to effectively detect SLA violations in real time.


%




In this paper, we ask: \emph{How can one design a real-time SLA monitoring system that dynamically allocates monitoring resources across diverse network slices according to their SLAs?}
We model the monitoring system as a \emph{closed control loop} that continuously observes the set of active slices, their SLAs, and relevant traffic metrics, and \emph{tunes} the data plane to collect the ``right'' amount of telemetry for each slice.
The first key challenge is to formalize the underlying resource-allocation problem in a way that is \emph{independent} of the specific data-plane mechanism used for telemetry collection.
This abstraction allows us to reason systematically about control-plane design and identify the minimal properties that a data-plane primitive must satisfy to enable slice- and SLA-aware closed-loop control of monitoring resources.

At a high level, our formulation introduces a ``tuning'' variable~$\knob$, defined per slice~$s$ and SLA metric~$m$, along with two functions, $E(\knob)$ and~$\Gamma(\knob)$, that capture the monitoring error and overhead, respectively, for a given choice of~$\knob$.
To make effective allocation decisions, the control plane must accurately estimate $E$ and~$\Gamma$, which depend on both the monitoring primitive and the characteristics of the active traffic in the network.

This leads to the second challenge: estimating these functions accurately is difficult since their behavior varies with both the primitive and traffic dynamics. For example, the bandwidth overhead of change-triggered selective INT -- which collects telemetry information only when metric fluctuations exceed a threshold \cite{chowdhury2021lint, sheng2021deltaint} -- depends on the variability of the monitored metric, which in turn depends on the characteristics of interacting traffic streams in a network (e.g., extent and timing of bursts). Similarly, in sketch-based monitoring primitives, the collision probability that affects measurement accuracy depends on the active traffic streams \cite{cormode05}. 

As such, in practice, $E$ and~$\Gamma$ must be continuously learned from the data-plane observations. This, however, is challenging since, at any moment, only one configuration of~$\knob$ values is deployed in the data plane, while the control loop needs estimates of $E$ and~$\Gamma$ across a \emph{range} of configurations to choose the most effective operating point.

Finally, the third challenge concerns the design of the data-plane primitive itself.
Because $E$ and~$\Gamma$ define a trade-off space between accuracy and overhead that the controller navigates per slice, the primitive must (1) expose a sufficiently wide range of operating points within this space, (2) support configurability at slice granularity, and (3) allow transitions between these operating points at runtime. As an extreme example, a primitive that always samples one out of ten packets would be unsuitable, as $E$ and $\Gamma$ would have the same value regardless of $\knob$, slice $s$, and SLA metric $m$.
%
We discuss these properties in detail in \S\ref{subsec:dp-implications}.

To demonstrate the feasibility of addressing these challenges and the benefits of slice-aware and SLA-aware closed-loop control, we design and implement an INT-based monitoring system, called \slicescope, that realizes this abstraction in practice.
We choose an INT-based data plane because slice monitoring in modern networks requires \emph{continuous, real-time} measurement of per-packet performance metrics, rather than aggregate statistics such as percentiles or averages. Moreover, SLA metrics like latency, jitter, and loss are increasingly defined \emph{end-to-end}, rather than per-hop~\cite{3gpp_22.261, 5gamericas2017, 5gamericas2021}.
%
INT enables network switches to add and update performance metrics to individual packets as they traverse the network \cite{int}, exposing per-packet end-to-end performance metrics. As such, it is well-suited for satisfying these requirements. 
INT, however, is known to incur non-negligible bandwidth overhead \cite{pint2020probabilistic}. As such, \slicescope opts for a change-triggered selective INT approach that enables trading off telemetry collection frequency (i.e., bandwidth overhead) against monitoring accuracy, while providing the flexibility and granularity required for slice-aware and SLA-aware closed-loop control.
While we focus on an INT-based implementation in this paper, the control framework is general and could be extended to other data-plane monitoring mechanisms, such as probabilistic INT or sketches, which we leave for future exploration.

\indsignpost{Summary of contributions} This paper makes the following contributions: 
\vspace{-0.1\baselineskip}
\begin{itemize}[leftmargin=*]

\item We formalize SLA-aware slice monitoring as a closed-loop control problem (\S\ref{sec:problem_formalization}), by introducing an abstract optimization framework parameterized by tuning knobs $\knob$ and trade-off functions $E(\knob)$ and $\Gamma(\knob)$. This formulation is independent of the specific telemetry primitive and identifies the minimal requirements a data-plane primitive must satisfy, which we term the Telemetry Primtive Contract (TPC).

\item We present \slicescope (\S\ref{sec:dataplane}), a system that realizes our framework by implementing the TPC with change-triggered INT. \slicescope introduces: (1) per-slice state tracking for slice-level control, (2) composable end-to-end semantics for bounded error guarantees, and (3) mathematical models that enable the control plane to estimate the $E(\knob)$-$\Gamma(\knob)$ trade-off for different knob configurations in the design space.

\item We demonstrate the feasibility and benefits of slice-aware and SLA-aware closed-loop control by implementing \slicescope on Intel Tofino switches and evaluating it on a testbed with emulated 5G traffic and through large-scale simulations with 300 slices exhibiting diverse SLA targets and traffic patterns (\S\ref{sec:evaluation}).
\end{itemize}

\indsignpost{Evaluation highlights} We demonstrate that \slicescope's adaptive control is essential: while change-triggered INT provides the foundation, static thresholds alone cause frequent SLA violations. Our control-plane reduces violations by up to $4\times$ for critical slices under comparable overhead. We also validate that change-triggered INT best realizes the TPC, outperforming alternatives like PINT~\cite{pint2020probabilistic} and LightGuardian~\cite{zhao2021lightguardian} for end-to-end tracking of slice SLAs, and perform microbenchmarks and test-bed evaluations.
\emph{This paper does not raise ethical issues.}

\indsignpost{Prior work} To the best of our knowledge, we are the first to formulate slice monitoring as a closed-loop control problem and define the \emph{Telemetry Primitive Contract (TPC)} that specifies the minimal data-plane requirements for such control. In \S\ref{subsec:why-delta-int}, we revisit prior telemetry mechanisms, and explain why change-triggered INT provides the best basis to realize the TPC in practice.




\section{Context: SLA Diversity in 5G and Beyond} \label{subsec:sla_diversity}

\indsignpost{SLA diversity in the real world: 5G slicing} 
While our work applies broadly to any network supporting slicing, we ground our discussion in 5G networks, where slicing is a core architectural primitive. 5G networks are expected to support dozens to hundreds of active slices \cite{atis-slicing, balasingam24, mittal-slicing}, each with distinct \glspl{SLA}. 
There are three broad slice categories: \gls{URLLC}, \gls{eMBB}, and \gls{mMTC} (see Table~\ref{table:workload-summary}). However, real-world deployments require more granularity. For example, within the broad URLLC category, discrete automation needs 10ms latency and 10Mbps throughput, while process automation needs 60ms latency and just 1Mbps throughput \cite{5gamericas2021}. 
Therefore, operators typically provision many slices, as opposed to just one for each broad category, each with distinct, fine-grained SLAs that must be continuously monitored in real-time for effective service assurance \cite{balasingam24,atlas-conext22}.
Other examples of slicing in modern networks include cloud providers providing service-specific connections with distinct performance guarantees (e.g., Azure ExpressRoute \cite{azure_expressroute}, Google Dedicated Interconnect \cite{gcp_dedicated_interconnect}), and wide-area networks providing differentiated services \cite{rfc9543}. 

\textbf{Why monitor per slice (as opposed to per flow)?} From a network operator's perspective, the natural monitoring unit is a \textit{slice} -- the level at which resources are provisioned and SLAs are defined. Slice-level monitoring offers two key advantages. 

\begin{itemize}[leftmargin=*]
\item \textit{Alignment with SLAs}: Slices are the natural unit of SLA definition and resource provisioning. Operators provision resources at slice granularity and commit to slice-level QoS guarantees, making slice-level monitoring the natural level for SLA verification and control decisions. The responsibility of managing \textit{individual flows within a slice} typically rests with the service provider using the slice.
\item \textit{Tractability}: Although a network may have millions of flows, the number of active slices is typically in the hundreds \cite{atis-slicing, balasingam24, mittal-slicing}. This makes dedicated and accurate (as opposed to approximate) per-slice state manageable in programmable switches. This also makes closed-loop optimization of resource allocation computationally tractable -- our evaluations (\S\ref{sec:eval-microbenchmarks}) show that joint optimization of monitoring resources over 300 slices completes in $<1$s, which enables real-time adaptation.
\end{itemize}

\section{Formalizing closed-loop SLA-aware slice monitoring} \label{sec:problem_formalization}


\begin{wrapfigure}{r}{0.4\columnwidth}
    \centering
    \vspace{-5pt}
    \includegraphics[width=0.4\columnwidth]{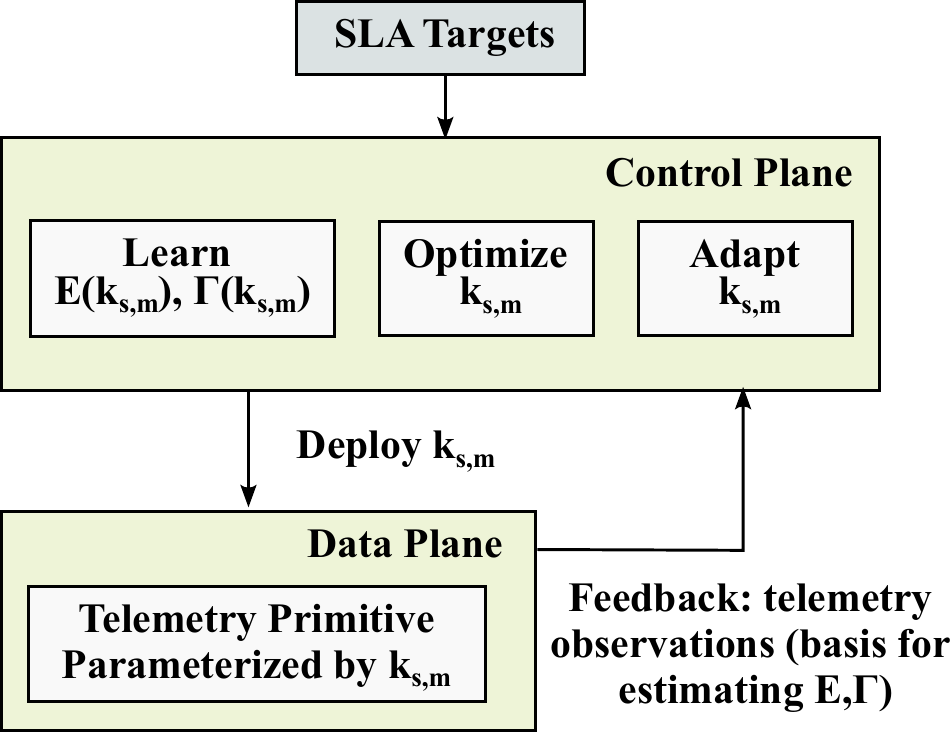}
    \vspace{-8pt}
    \caption{\textit{Closed-loop formulation of SLA-aware slice monitoring.} 
    The control plane continuously learns from telemetry observations that reflect slice-level performance, 
    estimates monitoring error $E(\knob)$ and overhead $\Gamma(\knob)$, and tunes per-slice knobs $\knob$ 
    of the data plane telemetry primitive to satisfy SLA objectives under resource constraints.}
    \label{fig:closed-loop control}
    \vspace{-5pt}
\end{wrapfigure}

As shown in figure~\ref{fig:closed-loop control}, we model the slice monitoring system as a closed control loop that continuously takes in the set of active slices, their SLAs, and relevant information about their traffic patterns and observed SLA metrics, and tunes the network data plane to collect the ``right'' amount of telemetry information for each slice. 
In this section, we describe how we formalize the resource allocation problem that the control loop must solve independent of the particular data-plane monitoring mechanism used to collect telemetry information (\S\ref{subsec:optimization}), how it can adapt to changing network conditions (\S\ref{subsec:update-strategy}), and its implications on the properties needed from the data-plane monitoring primitives (\S\ref{subsec:dp-implications}).

\subsection{Joint Multi-Slice Resource Allocation}
\label{subsec:optimization}

\indsignpost{SLA model and monitoring error tolerance} We model a slice SLA for a metric $m$ by its target value $\slaval$ (e.g., max latency of 10ms). Critical slices (e.g., URLLC) have stringent targets (e.g., $\slaval$=5ms) while non-critical slices (e.g., mMTC) have more relaxed targets (e.g., $\slaval$=40ms). Slice SLAs specify targets for end-to-end metrics but not acceptable \emph{monitoring} error, despite its acknowledged impact~\cite{rfc9543}.




\indsignpost{Slice monitoring as a joint resource allocation problem} 
Each slice–metric pair operates within its own accuracy–overhead trade-off space: allocating more monitoring resources (e.g., computational resources in network devices or telemetry bandwidth) yields higher measurement accuracy for that metric in that slice.
However, as we show in our evaluation (\S\ref{sec:eval-closing-loop}), independent per-slice optimization is suboptimal; \emph{joint optimization} enables coordinated trade-offs where slices with error-budget slack can accept reduced fidelity while near-constrained slices receive tighter monitoring.
As such, for a network with $N$ active slices and $M$ SLA metrics, the monitoring system must manage $N \times M$ ``instances'' of this trade-off space while balancing per-SLA monitoring accuracy with network-wide monitoring cost.

We model this as a constrained resource allocation problem, summarized in Table~\ref{tab:psto_optimization_summary}.
Specifically, our formulation captures the accuracy-overhead trade-off space in a unified manner with two functions, $E(x)$ and~$\Gamma(x)$, that are connected through \emph{the ``tuning'' parameter} $x$.  
This provides our formulation with a unified way to explore the $N \times M$ instances of the trade-off space using dedicated tuning variables for each slice and metric.
Concretely, we define a ``tuning'' variable~$\knob$ for each slice~$s$ and SLA metric~$m$, along with two functions, where $E(\knob)$ and~$\Gamma(\knob)$ capture the monitoring error and overhead, respectively, for a given choice of~$\knob$.
The goal is to select configurations $\{\knob\}$ that minimize system-wide cost -- sum of $\Gamma(\knob)$ across all slices and metrics -- while respecting all constraints with respect to SLA monitoring accuracy, i.e., $E(\knob) \leq \epsilon_{s,m}$. 

\begin{table*}[t!]
\setlength{\abovecaptionskip}{-15pt}
\centering
\small
\begin{minipage}[t]{0.4\textwidth}
\setlength{\abovecaptionskip}{3pt}
\centering
\begin{tabularx}{\linewidth}{>{\bfseries}p{1.6cm} X}
Objective & $\min_{K} \sum_{s,m} \lambda \, E(\knob) + (1 - \lambda) \, \Gamma(\knob)$ \\
\multicolumn{2}{l}{\textbf{Constraints}} \\
C1 & $\forall s,m: E(\knob) \leq \epsilon_{s,m}$ \\
C2 & $\forall s,m: \sum_c z_{s,m,c} = 1$ \\
C3 & $\forall s,m,c: z_{s,m,c} \in \{0, 1\}$ \\
C4 & $\forall s,m: K_{s,m} = \sum_k z_{s,m,c} \cdot K_{s,m,k}$ \\
\end{tabularx}
\caption*{\textbf{(a)} Optimization formulation}
\end{minipage}
\hfill
\begin{minipage}[t]{0.57\textwidth}
\setlength{\abovecaptionskip}{3pt}
\centering
\begin{tabularx}{\linewidth}{>{\bfseries}p{1.1cm} X}
Symbol & Description \\
\midrule
$E(\knob)$ & Expected monitoring error for slice $s$ and metric $m$ given $\knob$ as tuning parameter \\
$\Gamma(\knob)$ & Expected telemetry overhead for slice $s$ and metric $m$ given $\knob$ as tuning parameter\\
$\epsilon_{s,m}$ & SLA-specific monitoring error tolerance for slice $s$ and metric $m$\\
$\lambda$ & Trade-off between accuracy and overhead \\
$K_{s,m,c}$ & Candidate ``knob'' values \\
$z_{s,m,c}$ & Binary variable selecting $K_{s,m,c}$ \\
\end{tabularx}
\caption*{\textbf{(b)} Variable definitions}
\end{minipage}
\caption{SLA-aware multi-slice joint resource allocation }
\label{tab:psto_optimization_summary}
\end{table*}



\indsignpost{Objective: balancing accuracy and overhead among acceptable joint solutions} The objective is to minimize both expected monitoring error and telemetry overhead across all slices and metrics. 
Note that, as we discuss below, our formulation enforces that for each slice $s$ and metric $m$, the monitoring error $E(\knob)$ should not exceed the acceptable threshold $\epsilon_{s, m}$ through dedicated constraints. 
The objective includes a tunable weight~$\lambda$ that operators can adjust to express policy preferences.
Specifically, $\lambda$ allows operators to balance accuracy and overhead among all valid solutions: larger $\lambda$ favors more accurate solutions, whereas smaller $\lambda$ favors those with lower network-wide overhead. 
Moreover, including error in the objective (in addition to respecting per-slice tolerances) can prevent undesirable solutions, e.g., highly unbalanced ones that would otherwise just barely satisfy the monitoring requirements of critical slices.

\indsignpost{Accuracy and $\knob$ selection constraints} 
Each slice $s$ has an error tolerance $\epsilon_{s,m}$ for each metric $m$, with critical slices having lower tolerances than non-critical ones. 
Constraint C1 enforces that chosen $\knob$ values respect these tolerances. 
Moreover, our formulation selects $\knob$ from a discrete set of predefined candidate values.
Each $\knob$ represents a distinct operating point in the accuracy–overhead trade-off of the data-plane monitoring primitive.
While some primitives may conceptually support a continuous range of $\knob$ values (\S\ref{sec:dataplane}), real-world programmable switches (e.g., Intel Tofino~\cite{tofino}) typically implement them using fixed-point arithmetic with a configurable step size, yielding a finite set of realizable values.
For example, with an 8-bit fixed-point representation and step size $0.05$, $\knob$ can take on $256$ possible values between $0$ and $12.75$.

To generalize across different data-plane primitives and reflect these hardware constraints, we model the candidate values as $K_{s,m,c}$ and use binary variables $z_{s,m,c}$ to indicate which one is selected.
Constraints C2–C4 ensure that exactly one candidate value is chosen per slice–metric pair $(s,m)$.


\indsignpost{The challenge of learning trade-off functions}
To make effective allocation decisions, the control plane must accurately estimate the ``trade-off functions'' $E$ and~$\Gamma$, which depend on both the monitoring primitive and the characteristics of the active traffic in the network.
For example, the bandwidth overhead of change-triggered selective INT -- which collects telemetry information only when metric fluctuations exceed a threshold \cite{chowdhury2021lint, sheng2021deltaint} -- depends on the variability of the monitored metric, which in turn depends on the characteristics of interacting traffic streams in a network. Similarly, in sketch-based monitoring primitives, the collision probability that affects measurement accuracy depends on the active traffic streams \cite{cormode05}. As such, $E$ and~$\Gamma$ must be continuously learned from the data-plane observations.

What makes this challenging is that the control plane needs to construct models for predicting how \emph{arbitrary candidate values} of $\knob$ affect monitoring accuracy $E$ and overhead $\Gamma$ from \emph{limited observations} of a data plane that is configured with \emph{one set of concrete $\knob$ values} at that point in time.
We cannot simply try out different $\knob$ configurations. Trying suboptimal settings risks SLA violations during 
exploration. Moreover, with hundreds of slices, various metrics, and large parameter spaces, exhaustive exploration is impractical.
As such, the control plane must generalize from limited observations without requiring deployment of every candidate in the data plane.
As mentioned above, $E$ and $\Gamma$ also depend on the particular data-plane monitoring primitive used in the control loop. We describe how \slicescope, our realization of the closed-loop control with change-triggered INT as its data-plane primitive, addresses this challenge and estimates these functions in \S\ref{subsec:math_framework}.

%

\eat{
\signpost{The Challenge}  To optimize the tuning knob selection $\mathbf{\Delta}$, we must predict how other candidate $\mathbf{\Delta}$ values affect monitoring accuracy and overhead. Directly deploying candidates risks SLA violations, while exhaustive exploration is impractical with hundreds of slices.
}


\subsection{Adapting to Network Dynamics} \label{subsec:update-strategy}

The formulation in \S\ref{subsec:optimization} provides optimal $\knob$ values for each slice and metric. However, its effectiveness depends on $E$ and $\Gamma$, which can very over time with changing network conditions. 
As such, we divide time into \emph{epochs of $\tau$ seconds}, and poll the data plane at the beginning of each epoch to update the learned estimates of $E$ and $\Gamma$, and subsequently re-run the joint optimization (\S\ref{subsec:optimization}) to recompute $\knob$ values. The choice of epoch length $\tau$ is a trade-off: it must be short enough to capture network dynamics, yet long enough for the control plane to compute feasible thresholds.
We evaluate the impact of $\tau$ in \S\ref{sec:eval-microbenchmarks}.

\indsignpost{Update strategies} We need to update the $\knob$ values within bounded per-epoch computation time. What makes this challenging is that optimization may take a long time to converge, requiring approximate methods or fallbacks 
to ensure timely decisions (\S\ref{subsec:update-strategy}).
Our formulation is an Integer Linear Program (ILP), which are NP-complete in the general case.
As such, exact optimal solutions may take exponential time in the worst case.
To ensure robustness in the face of uncertain optimization times and enable timely adaptation, we use a two-tiered approach:

\textit{1. ILP with early stopping.} 
We limit solver runtime to a fraction of the epoch length $\tau$, to ensure that a solution is found in bounded time. Even if the global optimum is not reached, the best current solution is validated for feasibility before being deployed. Our evaluation shows that early stopping provides feasible solutions for $\tau$ as small as 5 seconds (\S\ref{sec:eval-microbenchmarks}). 

\textit{2. Greedy heuristic fallback.} 
If the ILP solver fails to find a feasible solution within the allotted time, we switch to a lightweight greedy heuristic. The heuristic iterates over slices in order of criticality (e.g., URLLC first).
For each slice-metric pair, it selects the set of candidate $\knob$ values (out of the discrete set of possible values) that satisfy that pair's monitoring error tolerance (based on $E$). If multiple candidates meet this criterion, the one with the lowest expected overhead is chosen.
When no candidate satisfies the tolerance, the heuristic defaults to the most conservative $\knob$ value (i.e., smallest error) to ensure SLA safety. 
This greedy approach prioritizes constraint satisfaction over the 
error-overhead trade-off: it ensures all $\epsilon_{s,m}$ bounds are met, then minimizes overhead among valid solutions, rather than jointly balancing the two objectives via $\lambda$. This preserves SLA guarantees while enabling timely decisions when full optimization is infeasible.
 
\subsection{Requirements from the data-plane telemetry primitive}
\label{subsec:dp-implications}

For the joint multi-slice optimization to be effective, we need a data plane primitive with particular capabilities in the closed control loop. Inspired by our general formulation of the problem in \S\ref{subsec:optimization} and \S\ref{subsec:update-strategy}, we derive a minimal set of such capabilities, which we call the \emph{Telemetry Primitive Contract (TPC)} and describe in this section.

\indsignpost{(R1) Per-slice per-metric tunability} The control-plane optimization relies on 
navigating the accuracy-overhead trade-off space -- through $E$ and~$\Gamma$ -- \emph{for each slice}.
As such, the data-plane telemetry primitive should expose \emph{per-slice, per-metric 
runtime knobs} that the control plane can adjust each epoch, 
enabling differentiation across slices with heterogeneous SLAs.
Moreover, these knobs should offer a wide range of operating points in this trade-off space, i.e., with a wide variety of values for $E$ and $\Gamma$.
A primitive that, say, always samples one out of ten packets would be unsuitable, as $E$ and $\Gamma$ would have the same value regardless of $\knob$, slice $s$, and SLA metric $m$.


\indsignpost{(R2) Run-time tunability} Not only should the data plane primitive offer a wide range of operating points in the accuracy-overhead trade-off space, it should allow transitions between them \emph{at runtime}.
That is, to be able to do closed-loop control, the control plane should be able to adjust data-plane knobs, and in turn, monitoring and accuracy for each slice-metric pair without having to recompile the data plane.



\textbf{(R3) Robust and composable E2E Semantics:} 
To allow the control plane to reason about slice-level SLA monitoring fidelity, we need a data-plane primitive with E2E telemetry semantics whose local measurements combine into a predictable end-to-end accuracy ($E$) and overhead ($\Gamma$) view (through strict analytical bounds or calibrated proxies), allowing the control plane to reason about slice-level SLA monitoring fidelity.
This is because modern SLAs are \emph{real-time} and increasingly defined \emph{end-to-end}, as opposed to per-hop \cite{3gpp_22.261, 5gamericas2017, 5gamericas2021}.
Tracking SLA metrics such as latency, jitter, and loss per-hop (as statistics or quantiles) is not effective for real-time compliance: without per-packet correlation across hops, we cannot tell whether observations at different hops affect the same or different packets.
For example, suppose a discrete automation application with an end-to-end latency \gls{SLA} of 15ms runs in a URLLC slice instance, and we find out 50\% of packets exceed 10ms at switch $A$, and 50\%  exceed 5ms at switch $B$. Without per-packet correlation, we cannot tell whether these are the \textit{same} packets accumulating delay across hops, or distinct packets encountering independent issues. 
Only the former indicates an \gls{SLA} violation.
As such, fine-grained (e.g., per-packet) end-to-end visibility is key to providing robust and composable E2E semantics that can be used as part of continuous closed-loop control.




\indsignpost{Summary} The TPC captures the minimal high-level requirements for a data plane telemetry primitive for SLA-aware network slice monitoring, without prescribing \textit{how} a primitive satisfies these requirements. In \slicescope (\S\ref{sec:dataplane}), we present one realization of the TPC using change-triggered selective INT; however, alternative primitives could satisfy it through different means.

\section{\slicescope: Realizing the Closed Loop with Change-Triggered INT} \label{sec:dataplane}


Building on \S\ref{sec:problem_formalization}, \slicescope realizes the real-time SLA-aware closed-loop control using a change-triggered INT-based telemetry in the data-plane. 
We first explain why change-triggered INT is uniquely well-suited for this framework (\S\ref{subsec:why-delta-int}), then develop the mathematical models for estimating $E(\knob)$ and $\Gamma(\knob)$ (\S\ref{subsec:math_framework}), describe the data plane implementation (\S\ref{subsec:data_structure}), and discuss additional mechanisms needed for practical deployment (\S\ref{subsec:corner_cases}).

\indsignpost{Generalizability} Although we instantiate the TPC abstraction using change-triggered selective INT, the abstraction itself is not tied to any specific primitive. Other telemetry mechanisms, such as future probabilistic or hybrid designs could satisfy the same contract if extended with equivalent slice-level control and e2e semantics. Together, our closed-loop control formulation (control-plane) and TPC (data-plane) provide a blueprint for any dynamic, SLA-aware monitoring system.


\subsection{Why change-triggered In-Band-Network Telemetry (INT)?} 
\label{subsec:why-delta-int}

\signpost{Existing telemetry mechanisms fall short} Section \S\ref{subsec:dp-implications} defined the abstract Telemetry Primitive Contract in the form of three requirements R1-R3. These requirements rule out several classes of monitoring approaches. \textit{Active probing} uses synthetic probes to collect telemetry for e2e paths. However, the measurements are for the probes themselves, not actual slice traffic. Probes can experience different treatment (e.g., ECMP path, queueing/scheduling, policing), so their telemetry is not representative of per-slice behavior; consequently, it fails to meet the TPC (R1–R3). \textit{Sketch-based} approaches (e.g., \cite{nitrosketch, zhao2021lightguardian}) allow recording traffic summaries and memory-efficient estimates of statistics such as top-k
heavy hitters or latency quantiles at a single switch. However, they cannot correlate observations across packets to reconstruct end-to-end behavior (R3); our evaluation (\S\ref{sec:eval-e2e-tracking}) shows this yields insufficient accuracy for SLA tracking. Moreover, they are typically not sufficiently tunable at run-time (R2).
\textit{Probabilistic sampling} \cite{pint2020probabilistic} aims at reducing overhead by instrumenting a subset of packets according to a configured sampling probability. While it is possible to adapt the sampling rate per-slice, probabilistically spreading telemetry across packets breaks the packet-path continuity needed for strict e2e semantics (R3). As we show in §\ref{sec:eval-e2e-tracking}, this loss of continuity leads to insufficient accuracy for SLA tracking.

\begin{table}[t!]
\setlength{\abovecaptionskip}{-10pt}

\centering
\scriptsize
\renewcommand{\arraystretch}{1.05}
\begin{tabularx}{\textwidth}{p{2.5cm} AYYYY}

& \multicolumn{2}{c}{\textbf{Data Plane Requirements (\S\ref{subsec:dp-implications})}} 
& \multicolumn{3}{c}{\textbf{SLA-aware Closed-Loop Control (\S\ref{subsec:optimization} \& \S\ref{subsec:update-strategy})}} \\
\cmidrule(lr){2-3} \cmidrule(lr){4-6}

& \textbf{\makecell{Slice-Level Run-\\Time Tunability \\ (R1-R2)}} 
& \textbf{\makecell{E2E \\ Semantics \\ (R3)}} 
& \textbf{\makecell{Trade-off\\Learning}} 
& \textbf{\makecell{Multi-Slice\\Optimization}}
& \textbf{\makecell{Runtime\\Adaptation}} \\

\midrule
\multicolumn{6}{@{}l}{\textbf{No e2e per-packet visibility}} \\
Active probing \cite{hohemberger2019orchestrating,bhamare2022intopt} 
& \xmark{}
& \xmark{} 
& \xmark{} 
& \xmark{} 
& \cmark{} \\

Sketches \cite{nitrosketch,mcsketch-infocom22,zhao2021lightguardian} 
& \xmark{} 
& \xmark{} 
& \xmark{} 
& \xmark{} 
& \xmark{} \\

Probabilistic \cite{pint2020probabilistic,tang2022orchestration}     
& \cmark{} 
& \xmark{}
& \xmark{}
& \xmark{} 
& \cmark{} \\

\midrule
\multicolumn{6}{@{}l}{\textbf{E2e visibility, but lacking control}} \\

Path-aggregated \cite{marques20intsight}             
& \xmark{}
& \cmark{} 
& \xmark{} 
& \xmark{}  
& \xmark{} \\

Change-triggered \cite{sheng2021deltaint,chowdhury2021lint}          
& \pmark{} 
& \pmark{} 
& \xmark{} 
& \xmark{} 
& \xmark{} \\

\midrule
\textbf{SliceScope (this work)} 
& \cmark{} 
& \cmark{} 
& \cmark{} 
& \cmark{} 
& \cmark{} \\
\bottomrule
\end{tabularx}
\caption{
Comparison of representative telemetry approaches against data-plane requirements (\S\ref{subsec:dp-implications}) and closed-loop control challenges (\S\ref{subsec:optimization} and \S\ref{subsec:update-strategy}). Only \slicescope{} satisfies all criteria. \cmark{} supported, \xmark{} not supported, \pmark{} partial support.
}
\label{tab:telemetry-comparison}
\end{table}

INT-based approaches provide the necessary per-packet, end-to-end correlation. \textit{Path-aggregated} INT \cite{marques20intsight} carries complete e2e metrics in fixed size packet headers, ensuring full visibility but at fixed overhead with no run-time per-slice control (R1 and R2). \textit{Change-triggered} INT \cite{sheng2021deltaint, chowdhury2021lint} is more promising: it inserts telemetry only when metrics deviate by more than threshold $\Delta$, naturally exposing a control knob. However, existing designs apply $\Delta$ globally and statically, and for one concrete metric, providing no concrete pathway to per-slice run-time adaptation and for a range of end-to-end common SLA metrics.
Thus our data plane requirements are only partially satisfied ($\sim$R1--R3). Moreover, the existing designs all lack any form of SLA-aware closed-loop control.

Table~\ref{tab:telemetry-comparison} summarizes these limitations against our data-plane requirements (R1–R3), as well as various components of our closed-loop control addressed by \slicescope and not by prior work.

\textbf{\slicescope} extends change-triggered INT to address these gaps. We make $\Delta$ a per-slice, per-metric knob $\Delta_{s,m}$, add mechanisms for composable end-to-end semantics, and enable the control plane to learn trade-offs, allocate thresholds across slices, and adapt them at runtime. This provides the missing link: a controllable primitive that enables the closed loop of learning, resource allocation, and adaptation required for real-time, SLA-aware monitoring.


\subsection{Mathematical Framework for Modeling $E$ and $\Gamma$} \label{subsec:math_framework}

Section \S\ref{subsec:optimization} identified that the control plane must estimate $E(\knob)$ and $\Gamma(\knob)$ from limited observations, a challenge whose solution depends on the specific 
telemetry primitive. 
5
In \slicescope, our telemetry primitive is changed-triggered INT. Specifically, we define per-slice, per-metric knob $\Delta_{s,m}$, and only add INT information to the packet 
only when metric $m$ for slice $s$ deviate by more than threshold $\Delta_{s,m}$ from previously collected value.
We set $\knob \equiv \dknob$, and develop the mathematical models that enable this estimation for change-triggered INT.
The core challenge is that only one threshold configuration, $\dknob$, is deployed in the data plane at any given time, yet the optimizer needs to evaluate many candidates to select the best one (\S\ref{subsec:optimization}). We address this by modeling the relationship between $\dknob$ and telemetry insertion decisions, which in turn, determines both monitoring accuracy and overhead.

\begin{figure*}[!ht]
\setlength{\abovecaptionskip}{-3pt}
\setlength{\belowcaptionskip}{-10pt}
    \centering
    \includegraphics[width=0.9\linewidth]{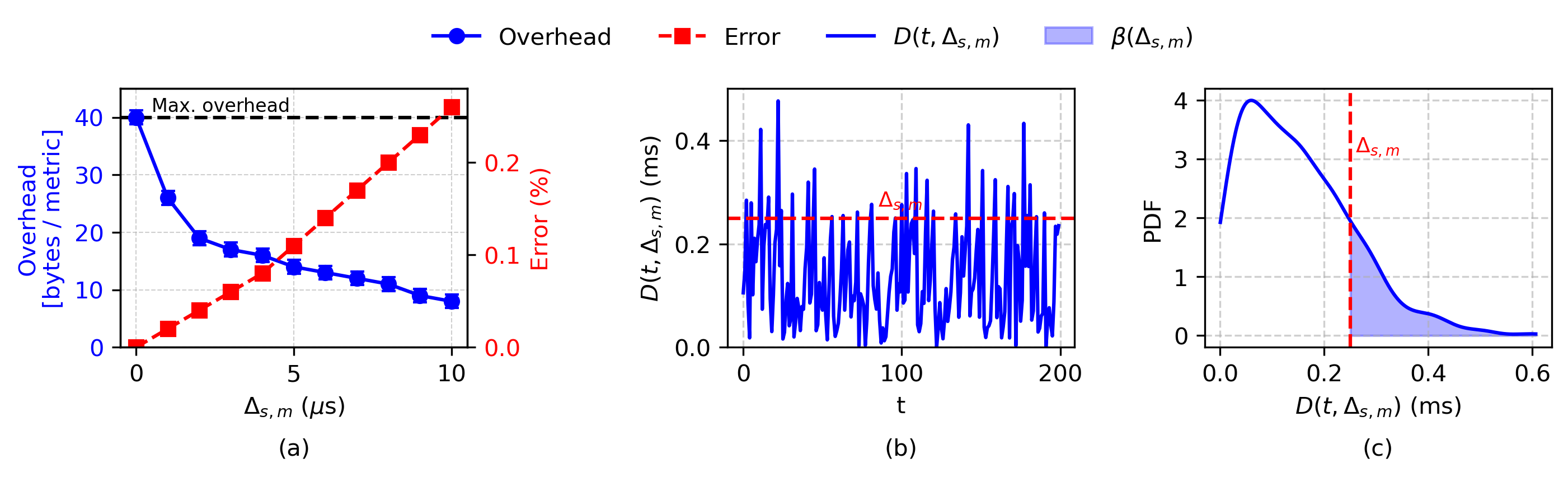}
    \caption{Analytical framework for change–driven telemetry insertion. \textbf{(a)} Accuracy–overhead trade-off as a function of the telemetry threshold $\Delta_{s,m}$: increasing $\Delta_{s,m}$ reduces telemetry overhead (blue) but increases monitoring error (red). 
    \textbf{(b)} Example trace of the simulated cumulative metric difference $D(t,\Delta_{s,m})$, which mimics how $\cknob$ accumulates 
in the data plane, and triggers telemetry insertion whenever it exceeds $\Delta_{s,m}$. 
    \textbf{(c)} Probability density of $D(t,\Delta_{s,m})$, where the shaded region corresponds to the insertion probability $\beta(\Delta_{s,m})$.}
    \label{fig:delta_tradeoff}
\end{figure*}

\indsignpost{Data plane operation (preview)}
Before developing the models, we briefly preview how change-triggered INT works (full description in \S\ref{subsec:data_structure}): Each switch maintains per-slice state, including the last reported metric value for each metric. When a packet arrives, the switch computes the accumulated difference $\delta_{s,m}$ between the current metric and the last reported value. If $\delta_{s,m}$ > $\dknob$, the switch inserts telemetry and 
resets $\delta_{s,m}$; otherwise, it skips insertion. This mechanism creates the knob: smaller $\dknob$ causes more insertions (higher accuracy, higher overhead), while larger $\dknob$ suppresses insertions, as shown in Figure~\ref{fig:delta_tradeoff}.

\smallskip

\indsignpost{Modeling telemetry insertion probability} Given this mechanism, the key is to quantify the relationship between $\dknob$ 
and insertion likelihood through the telemetry insertion probability:
$\beta({\Delta_{s, m}}) = \mathbb{P}(\delta_{s,m} > \Delta_{s, m})$. However, directly modeling $\cknob$ creates a circular dependency: circular dependency: the accumulated difference $\cknob$ depends on the last reported state, which itself depends on when telemetry was last inserted (i.e., when $\cknob$ previously exceeded the threshold currently deployed). This circular dependency makes direct modeling of $\beta({\Delta_{s, m}})$ impractical.

\indsignpost{Learning underlying packet differences} Our key insight is that while we cannot directly model $\cknob$, we can instead work with the underlying packet-to-packet differences that generate these deviations. Specifically, we learn the distribution of successive metric differences $d(t) = E_{curr}^t - E_{last}^t$ from the switch data plane, which capture how the metric changes between consecutive packets at time $t$. Unlike $\cknob$, these packet differences are independent of the chosen threshold. Using this learned distribution, we can simulate, in the control plane, how different threshold values would affect telemetry insertion. We do this by computing the cumulative sum $D(t, {\Delta_{s, m}}) = |\sum d(t)|$, resetting it whenever it reaches $\Delta_{s, m}$. This mimics exactly how $\cknob$ accumulates and resets in the data plane. Thus, the cumulative sum $D$ effectively serves as a proxy for $\cknob$. This simulation enables us to construct the probability function for arbitrary thresholds. Figure~\ref{fig:delta_tradeoff} illustrates this process, showing how the telemetry insertion probability $\beta(\Delta_{s,m})$ is computed as area under the probability density of $D(t, {\Delta_{s, m}})$.

Having established how different thresholds determine telemetry insertion probabilities, we next examine how these probabilities affect monitoring accuracy along a path. Skipped insertions reduce overhead, but also allow errors to accumulate across switches. Quantifying and bounding this accumulation is critical to ensure SLA error tolerances are respected.

\indsignpost{Modeling error accumulation} Consider packets from slice $s$ traversing a path $\mathcal{P}$. Let $\eta_i$ denote the absolute error between the true and measured value of metric $m$ at switch $i$. At each hop $i{-}1$, telemetry is inserted with probability $\beta_{i-1}(\Delta_{s,m})$. The expected error therefore evolves as:
\[
\E[\eta_i] = \E[\eta_{i-1}] + \big(1-\beta_{i-1}(\Delta_{s,m})\big)\cdot \Delta_{s,m}.
\]
Intuitively, when telemetry is inserted (probability $\beta_{i-1}$), the current switch simply inherits the previous error $\eta_{i-1}$. On the other hand, when it is skipped (probability $1-\beta_{i-1}$), the error grows by up to $\Delta_{s,m}$ in addition to $\eta_{i-1}$.


\indsignpost{Bounding error growth}
A naive worst-case argument assumes every hop skips insertion, yielding an upper bound of $(|\mathcal{P}|-1)\cdot \Delta_{s,m}$ (one potential $\Delta_{s,m}$ contribution per hop).
However, insertions occur with probability $\beta_j(\Delta_{s,m})$ and reset the error. Unrolling the recurrence above (by induction on path length) gives: 
$$E(\knob) = E(\Delta_{s,m}) = \mathbb{E}[\eta]_{UB}
= \Big( (|\mathcal{P}| - 1) - \sum_{j=1}^{|\mathcal{P}| - 1} \beta_{j}(\Delta_{s,m}) \Big)\, \Delta_{s,m}$$

This is a \emph{$\beta$-discounted} (tighter) upper bound: it matches the naive bound exactly when $\beta_j=0$ for all hops (no insertions), and is strictly smaller whenever any $\beta_j>0$. Hence it avoids the overprovisioning implied by the naive bound.
This is what we use for the monitoring error $E(\knob)$ in the closed-loop control's optimization problem in \S\ref{subsec:optimization}.


\indsignpost{Estimating per-packet monitoring overhead}  
Having established how errors accumulate, we next quantify the cost of telemetry in terms of expected header overhead. 
Each monitored packet consists of: (i) a per-packet shim and bitmap ($b_0$ bits), 
(ii) per-hop metadata that is always present ($b_h$ bits per hop), 
and (iii) metric payloads ($b$ bits each) included only when $\cknob >\Delta_{s,m}$, i.e., with probability $\beta_j(\Delta_{s,m})$ at hop $j$ (see \S\ref{subsec:data_structure} for details). The expected per-packet overhead for metric $m$ on slice $s$ is therefore
$$
\Gamma(\knob) = \Gamma(\Delta_{s,m}) = \E[\gamma] = (b_0 + b_h)\cdot |\mathcal{P}| \;+\; b \sum_{j=1}^{|\mathcal{P}|} \beta_{j}(\Delta_{s,m}).
$$
The first term reflects the fixed cost that scales linearly with path length, while the second term captures the conditional metric payloads and grows in proportion  to the insertion probabilities. 
We use this as the monitoring overhead $\Gamma(\knob)$ in the closed-loop optimization (\S\ref{subsec:optimization}).

\subsection{Telemetry Data Structures and Operations}
\label{subsec:data_structure}


Having developed models for $E(\dknob)$ and $\Gamma(\dknob)$ (\S\ref{subsec:math_framework}), we now describe how the 
data plane realizes change-triggered INT and extends it to satisfy the Telemetry Primitive Contract from \S\ref{subsec:dp-implications}.

Existing change-triggered INT designs apply thresholds globally and lack mechanisms for composable end-to-end semantics. \slicescope addresses these gaps through: (1) per-slice state tracking that enables slice-specific thresholds $\dknob$ (R1, R2), and (2) a two-part telemetry header with conditional insertion that provides bounded end-to-end error and overhead guarantees (R3).

We first describe the data structures, then walk through the 
four-step processing algorithm with examples for latency, jitter, and loss. Deployment mechanics (e.g., MTU headroom, multipath disambiguation, recovery) are addressed in \S\ref{subsec:corner_cases}.

\begin{figure}[!ht]
\setlength{\belowcaptionskip}{-12pt}
    \centering
    \begin{minipage}[t]{0.45\columnwidth}
        \centering
        \includegraphics[width=0.7\linewidth]{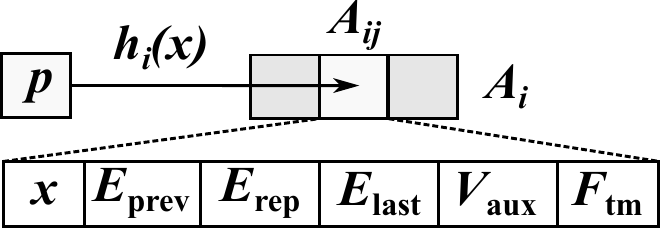}
        \caption{Per-slice telemetry state stored in each bucket $A_{ij}$ of the bucket arrays. 
        Each bucket maintains the slice key $x$, e2e estimates ($E_{prev},E_{rep},E_{last}$), 
        auxiliary state $V_{aux}$ (e.g., counters for packet loss), and a miss flag for recovery.}
        \label{subfig:hash_table}
    \end{minipage}
    \hfill
    \begin{minipage}[t]{0.53\columnwidth}
        \centering
        \includegraphics[width=0.7\linewidth]{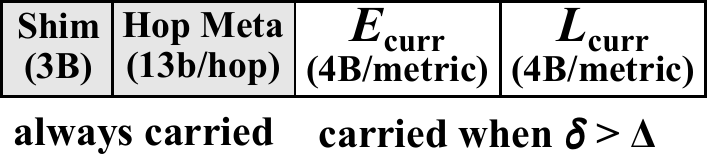}
        \caption{Telemetry header format. The fixed part (always present) includes a shim and per-hop 
        metadata. The conditional part (inserted only when $\cknob >\Delta_{s,m}$) 
        carries updated e2e metric values $E_{curr}$ and auxiliary state $V_{aux}$ for computing 
        metrics such as jitter or packet loss.}
        \label{subfig:telemetry_header}
    \end{minipage}
\end{figure}

\indsignpost{Per-slice state tracking} Each switch maintains $d$ bucket arrays, each with $w$ buckets, to track telemetry state for active slices (Figure~\ref{subfig:hash_table}). On packet arrival, a key $x$ including the slice ID is extracted from the packet headers. Next, for each bucket array $A_i$, an independent hash function $h_i(x)$ is used to map this packet to a bucket $A_{ij}$  (Figure~\ref{subfig:hash_table}). Each bucket stores: (1) the key $x$, (2) three end-to-end values for a given metric \textemdash{} $E_{prev}$ (estimate from previous hops), $E_{rep}$ (last reported), and $E_{last}$ (value from most recent packet), (3) auxiliary metrics $V_{aux}$ needed for computation of some metrics (e.g., estimated flow size for computing packet loss), and (4) a 1-bit table miss flag $F_{tm}$ to help recovery when downstream telemetry state is lost due to hash collisions. This per-slice state tracking enables slice-level customization of telemetry behavior (R1), as each slice can use its own threshold $\Delta_{s,m}$ to determine whether telemetry is inserted or skipped, at runtime (R2).

\signpost{Telemetry header} The telemetry header (Figure~\ref{subfig:telemetry_header}) has two parts: a \textit{fixed part} which is always carried, and a \textit{conditional part} only carried when significant changes occur (only when $\delta > \Delta_{s, m}$). The fixed part consists of (1) a telemetry shim with a bitmap indicating which selective telemetry fields are present, and (2) per-hop metadata containing a 10-bit node ID and three 1-bit anomaly flags. The conditional part consists of (1) current end-to-end values $E_{curr}$ for the slice metric, and (2) any auxiliary metrics $V_{aux}$ needed to compute that slice metric. For example, to compute \gls{E2E} packet loss, each switch appends the number of packets it forwarded for that slice ($V_{aux}$), allowing the next hop to estimate loss by comparing it with its own packet count.

Next, we show how the telemetry header and per-slice state are used to track per-slice \glspl{SLA} in an end-to-end per-packet manner with bounded error (R3).
\slicescope{}'s data plane algorithm runs in the switch egress pipeline, and processes incoming packets in the following steps: 

\signpost{Step 1 (Bucket array lookup)} The switch computes the key $x$ based on the packet headers, and looks it up in the bucket arrays. If a match is found, the switch retrieves the relevant telemetry state from previous packets, including $E_{prev}$ (estimate from previous hop) and $E_{rep}$ (last reported). 

\signpost{Step 2 (Metric computation)} The metric computation step involves three key operations:
\begin{itemize}[topsep=3pt,leftmargin=*]
\item \emph{Compute the current per-hop value  ($L_{curr}$):} This can be derived using either internal device state (e.g., hop latency) or auxiliary metrics (e.g., hop loss). Note that this value is different from the \gls{E2E} metric values $E_{curr}$, which are aggregated across the slice path.

\item \emph{Update the previous hop estimate of the e2e metric ($E_{prev}$):} If the packet carries a new update from the previous hop, it will be used and stored as the new $E_{prev}$. 
\item \emph{Compute the current e2e value  ($E_{curr}$):} This is done by combining the (potentially updated) previous hop estimate $E_{prev}$ with $L_{curr}$ using a simple aggregation operation (e.g., sum or max). 
\end{itemize}

\signpost{Step 3 (Selective telemetry insertion)} The switch compares the current \gls{E2E} value $E_{curr}$ with the last reported value $E_{rep}$. If the difference $\delta$ exceeds a given threshold $\Delta_{s, m}$ for that slice's metric, it is considered a significant change. The switch then inserts the current \gls{E2E} value $E_{curr}$ in the e2e metrics part of the packet header, and, when required, auxiliary information $V_{aux}$ to support downstream metric computation. For example, to compute \gls{E2E} packet loss across multiple hops, each switch inserts the number of packets it forwarded ($V_{aux}$), enabling the next hop to detect missing packets and estimate per-hop loss.

\signpost{Step 4 (State update)} If telemetry data is added to the packet, the switch updates the last reported \gls{E2E} value $E_{rep}$ in the corresponding bucket. We also update $E_{last}$ to be $E_{curr}$.

\begin{table*}[!ht]
\setlength{\abovecaptionskip}{-10pt}
\centering
\small
\renewcommand{\arraystretch}{1.2}
\begin{tabular}{|c|c|c|c|c|c|c|c|}
\hline
\textbf{Pkt} & \textbf{Sw} & \textbf{State Before} & $L_{curr}$ & $E_{curr}$ & $|\delta|$ vs $\Delta$ & \textbf{Action} & \textbf{State After} \\
\hline
p1 & $s_1$ & $E_{prev}{=}0$, $E_{rep}{=}0$ & 5 & $0 + 5 = 5$ & $|5{-}0| = 5 \ge 2$ & Add telemetry (5) & $E_{rep} \leftarrow 5$ \\
p1 & $s_2$ & $E_{prev}{=}5$, $E_{rep}{=}0$ & 6 & $5 + 6 = 11$ & $|11{-}0| = 11 \ge 2$ & Add telemetry (11) & $E_{rep} \leftarrow 11$ \\
p2 & $s_1$ & $E_{prev}{=}0$, $E_{rep}{=}5$ & 4 & $0 + 4 = 4$ & $|4{-}5| = 1 < 2$ & Skip telemetry & No change \\
p2 & $s_2$ & $E_{prev}{=}5$, $E_{rep}{=}11$ & 8 & $5 + 8 = 13$ & $|13{-}11| = 2 \ge 2$ & Add telemetry (13) & $E_{rep} \leftarrow 13$ \\
\hline
\end{tabular}
\caption{Walkthrough on a 2-hop path ($s_1 \!\rightarrow\! s_2$): for each packet,
the switch computes the local observation $L_{curr}$, updates the end-to-end estimate
$E_{curr}$, compares $\delta{=}|E_{curr}-E_{rep}|$ to $\Delta_{s,m}$, and inserts telemetry
only on threshold crossings.}
\label{tab:telemetry-example}
\end{table*}

\signpost{Walkthrough Example} Consider the simple case of a linear two-hop path with switches $s_1$ and $s_2$, where $\Delta = 2$ for both switches. Table~\ref{tab:telemetry-example} walks through how two packets ($p_1$, $p_2$) are processed at each hop. For packet $p_1$: At $s_1$, the local observation is $L_{curr} = 5$, giving an end-to-end estimate $E_{curr} = 0 + 5 = 5$. Since $\delta = |5 - 0| = 5 \geq 2$, telemetry is added and $E_{rep}$ updates to 5. At $s_2$, the local observation is 6, updating $E_{curr}$ to $5 + 6 = 11$. Since $\delta = |11 - 0| = 11 \geq 2$, telemetry is again inserted.
For packet $p_2$: At $s_1$, $L_{curr} = 4$ gives $E_{curr} = 0 + 4 = 4$, but $\delta = |4 - 5| = 1 < 2$, so no telemetry is added. At $s_2$, $L_{curr} = 8$ updates $E_{curr}$ to $5 + 8 = 13$. Since $\delta = |13 - 11| = 2 \geq 2$, telemetry is inserted.

The actual end-to-end latency for packet $p_2$ is $4 + 8 = 12$, but the reported estimate is 13 (using the stale $E_{rep} = 5$ from $s_1$). This creates an error of $|12 - 13| = 1$ at switch $s_1$, which is bounded by $\Delta_{s,m} = 2$. This example shows how selective insertion reduces overhead (one skipped report) while maintaining per-switch error bounds. Section~\ref{subsec:math_framework} describes how these per-switch bounds can be combined to derive end-to-end error guarantees.

\smallskip

\signpost{Metric-specific instantiations} The same four-step logic generalizes to different SLA metrics by adapting the definition of the local observation $L_{curr}$ and any auxiliary state $V_{aux}$. For example:

\begin{itemize}[leftmargin=*]
\item \textbf{End-to-end latency.} 
Each switch computes $L_{curr}$ as the difference between its egress and ingress 
timestamps ($t_{eg}-t_{ig}$). The end-to-end estimate is updated additively as 
$E_{curr}=E_{prev}+L_{curr}$, requiring no auxiliary state. 

\item \textbf{End-to-end jitter.} 
Here $L_{curr}$ is the change in per-hop latency relative to the previous packet. 
This requires storing the prior packet’s latency as $V_{aux}$; the switch then 
computes $L_{curr}=L_{curr}^{latency}-V_{aux}$ and updates the end-to-end jitter 
estimate additively. 

\item \textbf{End-to-end packet loss.} 
Each switch maintains a per-slice packet counter and, when telemetry is inserted, 
exports its current counter value $V_{aux}$ in the packet header 
(cf. Fig.~\ref{subfig:telemetry_header}). The downstream hop computes 
$L_{curr}$ as the difference between the upstream counter and its own local count, 
and aggregates this into the cumulative end-to-end loss estimate 
($E_{curr}=E_{prev}+L_{curr}$). 
\end{itemize}

In all cases, the dataplane executes the same four-step loop; only the definitions of $L_{curr}$ and $V_{aux}$ differ by metric. These examples show how \slicescope's threshold-based logic enables 
\emph{end-to-end}, per-slice monitoring of \emph{diverse SLA metrics} (latency, jitter, loss) with bounded error. We next address deployment aspects that preserve these semantics in practice (\S\ref{subsec:corner_cases}).

\subsection{Deployment Considerations: Ensuring Boundedness \& Stability} 
\label{subsec:corner_cases}

The data plane design (\S\ref{subsec:data_structure}) enables adaptive per-slice telemetry with bounded error. However, making this \textit{practical} requires addressing additional challenges: (1) ensuring telemetry headers fit within MTU constraints, (2) maintaining correctness under multipath routing, and (3) recovering from missing telemetry state.
\slicescope has mechanisms to address these challenges, which, together, make its data plane \emph{deployable in practice}. 

\indsignpost{1. Bounded telemetry header} A key deployability challenge is ensuring 
\emph{predictable header size}. Because sources cannot know in advance whether 
a packet will carry conditional telemetry, operators need a bounded 
worst-case header size to avoid fragmentation. \slicescope{}'s telemetry header 
(Figure~\ref{subfig:telemetry_header}) is designed with this in mind: it has 
(i) a fixed part that grows with the hop count $H$, and (ii) conditional 
fields that grow with the number of reported metrics $M$ but are independent 
of $H$.  The fixed part consists of a 3\,B shim plus 13\,bits per hop 
(10b node ID + 3b flags), rounded to bytes. The conditional part, carried 
only when $\cknob >\Delta_{s,m}$, adds at most 8\,B per metric 
(4\,B $E_{\text{curr}}$ and, when needed, 4\,B $V_{\text{aux}}$). The maximum 
header size is therefore $
3 \;+\; \left\lceil \tfrac{13H}{8}\right\rceil \;+\; 8M \;\; $ bytes.

Crucially, the \textit{heavier conditional fields do not scale with path length, 
unlike in traditional INT}. Even for three metrics (latency, jitter, loss) 
and $H{=}8$ hops, the total is only $3+13+24=40$\,B. For longer paths 
($H{=}16$) this grows to just $53$\,B. In practice, operator 5G transport 
networks are typically engineered with fewer than $\sim$15 hops~\cite{hanli2017itu}, 
so provisioning a small constant headroom (e.g., 64\,B via TCP MSS/MTU settings) ensures no fragmentation even in worst-case conditions.

\indsignpost{2. Multi-path support} 
Slice traffic often traverses multiple paths, i.e., distinct sequences of 
switches between the same endpoints. Aggregating telemetry state across 
paths can obscure performance variations and compromise accuracy. 
For example, suppose packets $p_1$ and $p_3$ of slice $s$ traverse 
$s_1 \!\rightarrow\! s_2 \!\rightarrow\! s_3$, while $p_2$ takes 
$s_1 \!\rightarrow\! s_4 \!\rightarrow\! s_3$. 
If switch $s_3$ maintains a single telemetry state for slice $s$, then $p_2$ 
updates $E_{rep}$ based on its higher-latency path. 
When $p_3$ arrives via the shorter path, its $\delta$ will be computed against 
$p_2$’s state, potentially suppressing telemetry that should have been reported.

To avoid this, \slicescope keys its telemetry state not only by slice ID but 
also by \emph{path identifier} and output port. 
The path ID distinguishes between upstream routes that converge at the 
same switch, while the output port disambiguates diverging downstream paths. 
To disambiguate paths without incurring large header overheads, 
\slicescope uses a compact 16-bit path identifier, pre-assigned by the control plane based on known slice routes. 
The path ID is constant-size and avoids carrying an explicit list of switch IDs. 
In practice, a 16-bit field suffices: with $\sim$300 slices and a few disjoint paths per slice ($\sim$1000 paths total), 
the probability of collision remains below 0.05\%.\footnote{By the birthday bound with $M=2^{16}$ identifiers and $N \approx 1000$ paths: 
$P_{\text{collision}} \approx 1 - e^{-N(N-1)/(2M)} \approx 0.0005$.}

\indsignpost{3. Recovering from state loss} 
Finally, state in the bucket arrays may be lost due to collisions 
or selective INT suppressing upstream updates. 
If a packet from slice $k$ arrives at $s_3$ but no valid $E_{prev}$ is found, 
accurate E2E computation would be impossible. 
\slicescope recovers by using the programmable packet replication engine to 
send a \emph{table miss notification} upstream. 
Upon receiving such a notification, the upstream switch sets a “table miss” bit 
for slice $k$, forcing the next packet to carry full telemetry regardless of 
its $\delta$ relative to $\Delta_k$. 
If that upstream switch has also lost state, the notification propagates 
recursively until it reaches the ingress of the slice path, where the state 
can be reinitialized. 
This ensures that downstream visibility is restored quickly after any loss. 
We evaluate the additional overhead of this mechanism in \S\ref{sec:eval-microbenchmarks}.

\smallskip

\indsignpost{Summary} \slicescope{} implements the closed-loop framework from \S\ref{sec:problem_formalization} using change-triggered INT with per-slice thresholds $\dknob$. We developed mathematical models to estimate monitoring accuracy $E(\dknob)$ and overhead $\Gamma(\dknob)$ (\S\ref{subsec:math_framework}), designed a data plane that provides slice-level control and end-to-end semantics (\S\ref{subsec:data_structure}), and developed mechanisms to enable practical deployment of \slicescope (\S\ref{subsec:corner_cases}). We now evaluate whether this design achieves the framework's goals of SLA-aware monitoring with controlled overhead.

\section{Evaluation} \label{sec:evaluation}

We demonstrate the feasibility and benefits of slice-aware and SLA-aware closed-loop control by evaluating \slicescope{} as its concrete realization along three main aspects:

\begin{itemize}[leftmargin=*]

\item \signpost{The benefit of per-slice closed-loop control (\S\ref{sec:eval-closing-loop})} Unlike prior work (e.g., \cite{sheng2021deltaint, chowdhury2021lint}) that relies on a single, static threshold, \slicescope{} continuously repositions each slice within the monitoring accuracy–overhead trade-off space.  
We demonstrate the benefit of this dynamic, SLA-aware adaptation by comparing against two baselines: a \emph{static slice-agnostic} control plane (same theshold for all slices) and a \emph{static slice-aware} control plane (different thresholds for each slice type, fixed over time).  

\item \signpost{Choice of data-plane monitoring primitive (\S\ref{sec:eval-e2e-tracking})} In \S\ref{subsec:why-delta-int}, we discuss why we choose change-triggered selective \gls{INT} over per-hop sketches or probabilistic selective \gls{INT}. In this section, we further justify this choice by empirically comparing \slicescope \textit{against LightGuardian \cite{zhao2021lightguardian} and PINT \cite{pint2020probabilistic}} in their ability to provide \emph{accurate end-to-end} tracking of per-slice \gls{SLA} metrics.

\item \signpost{Microbenchmarks (\S\ref{sec:eval-microbenchmarks}) and testbed evaluation (\S\ref{sec:eval-testbed})} We evaluate the impact of the number of slices on the control plane's decision time, the effect of epoch length on monitoring performance, and the impact of bucket array size on reporting frequency and other resource overheads. We also demonstrate \slicescope's effectiveness and hardware resource overheads over a testbed with Intel Tofino switches and emulated 5G traffic.

\end{itemize}

\subsection{Experimental Setup, Workloads, and Implementation}
\label{sec:eval-setup}

While our work applies broadly to any network supporting slicing, we have grounded our motivating example and experiments in 5G networks, where slicing is a core architectural primitive and networks are expected to support hundreds of active slices \cite{atis-slicing, balasingam24, mittal-slicing}, each with distinct \glspl{SLA} (\S\ref{subsec:sla_diversity}).

\indsignpost{5G-Inspired topology and slices} Our simulation topology is based on Telecom Italia's regional transport network~\cite{5g-metro-haul}: with 25 Gbps access links, 40 Gbps aggregation links, and a 100 Gbps core. We use 300 network slices of three main types: URLLC, eMBB, and mMTC (Table~\ref{table:workload-summary}). Network switches do weighted round-robin scheduling with weights decided based on the criticality of the main slice type and 22MB of buffer per-port \cite{buffers}. 
See \S\ref{sec:eval-microbenchmarks} for \slicescope's default parameters.


\begin{table*}[!ht]
\setlength{\abovecaptionskip}{-14pt}
    \centering
    \scriptsize
    \begin{tabularx}{\textwidth}{>{\bfseries}l *{5}{X} *{3}{c}}
\multirow{2}{*}{\shortstack{Main \\ Slice \\ Type}} 
& \multicolumn{5}{c}{\shortstack{\textbf{Range of performance metrics for \gls{SLA}} \\ \textbf{purposes and expected traffic pattern \cite{3gpp_22.261, 5gamericas2017, 5gamericas2021}}}} & \multicolumn{3}{c}{\shortstack{\textbf{Composition of Evaluated Workloads} \\ (\% of slice instances of each main type) }} \\ \cmidrule(lr){2-6} \cmidrule(lr){7-9}
         & \shortstack{E2E Lat.\\ (ms)}  & \shortstack{Loss \\ (\%)} & \shortstack{Jitter \\ (ms)} & \shortstack{Pkt. Size \\ (B)} & \shortstack{BW \\ (Mbps/user)} & \shortstack{More small \\ pkts (SP)} &
         \shortstack{Balanced \\ (BAL)} &
         \shortstack{More large \\ pkts (LP)}
         \\ \midrule
uRLLC    & 1-5       & <0.001  & <1        & 20-250    & 1-10               
& 60\% & 33\% & 20\%
\\
eMBB     & 10-50       & <1      & 5-30      & 1K-1.5K  & 15-50              
& 20\% & 33\% & 60\%
\\
mMTC     & 50-100    & 1-10    & 50-100    & 20-125  & 0.001-0.1          
& 20\% & 33\% & 20\%
\\
\midrule
\multicolumn{6}{r}{\textbf{Total \# Slices in Workload}} & 300 & 300 & 300 \\
\end{tabularx}
\caption{Summary of slice types and workloads used in the evaluation. Each slice type reflects a class of applications with similar SLA characteristics; many instances of each slice type are generated, each with its own SLA within the specified range. Bandwidth values refer to typical per-user usage; total slice bandwidth is computed as: (App. BW) × (\#users per slice), which is 3–10 (uRLLC), 10–20 (eMBB), and $\sim$10,000 (mMTC). Representative applications include industrial automation (uRLLC), UHD video streaming (eMBB), and sensor telemetry (mMTC). Details on workload generation are discussed in \textbf{\S\ref{sec:eval-setup}}.}
        \label{table:workload-summary}
\end{table*}

\indsignpost{Workloads and \glspl{SLA}} 
The impact of monitoring overhead varies with packet size -- small packets are more affected by \gls{INT} headers. 
Moreover, the ``right'' point in the per-slice accuracy-overhead trade-off depends on the \glspl{SLA} of active slices -- fewer critical slices means the control plane can use lower $\Delta$ values more freely, as less-critical slices require fewer \gls{INT} reports.
%
%
As such, we generate three different workloads with different compositions of slice types and packet sizes by changing the fraction of active slice instances that are of each of three main slice types.

Each main slice type has a distinct range of expected performance metrics and packet sizes in the 5G literature 
 \cite{3gpp_22.261, 5gamericas2017, 5gamericas2021}. For example, URLLC and mMTC slices have much smaller packets than eMBB, and URLLC has stricter latency requirements than eMBB, which in turn is stricter than mMTC (Table~\ref{table:workload-summary}).
Accordingly, the compositions of our three workloads are: (1) \emph{More small (and highly-critical) packets (SP)} with 60\% of slices of the URLLC type and 20\% for each of eMBB and mMTC, (2) \emph{Balanced (BAL)} with equal fraction of slices from each main type, and (3) \emph{More large (and less highly-critical) packets (LP)} with 60\% eMBB, 20\% URLLC, and 20\% mMTC.
For each workload, we generate traffic by creating the set of active slices, each with randomly assigned \glspl{SLA}, flow counts, packet sizes, and per-flow rates from expected ranges of these values for the corresponding main slice type. 
%
We set the monitoring error tolerance to $5\%$ of each metric's \gls{SLA} value.
 %


\indsignpost{Implementation} We have implemented \slicescope's data-plane pipeline (\S\ref{sec:dataplane}) using P4 \cite{p4} ($\approx2K$ lines of code) in the egress pipeline of an Intel Tofino switch \cite{tofino}. 
The telemetry data is carried in an INT shim header and metadata stack between the TCP/UDP headers and the payload. In our Tofino implementation, we added 3 pad bits to the 13-bit per-hop metadata header to meet byte-alignment requirements.
%
%
We use the \gls{TEID} from the \gls{GTP} protocol used in 5G as the \texttt{slice\_id}. In networks using \gls{SRv6}, \texttt{slice\_id} can be embedded in the SRv6 header \cite{ietf-slicing}. In other networks, such as MPLS, the MPLS labels, added by the ingress or provider edge router can be used instead.
\slicescope's control plane is implemented in Python, using the proprietary Gurobi library \cite{gurobi} to solve the ILP. 
To estimate the telemetry insertion probabilities $\beta(\Delta_{s,m})$, we fit a Laplace distribution to packet differences $d(t)$ after learning the distribution parameters from the data plane. The control plane runs on a server
equipped with an Intel i9 5.7 GHz processor and 32 GB RAM. 
For large-scale experiments, we developed a discrete-event simulator using SimPy \cite{simpy} that replicates the behavior of our P4 implementation. We will make our implementation available with the published paper.

\subsection{The Benefit of Per-Slice Closed-Loop Control }
\label{sec:eval-closing-loop}

\slicescope's control plane continuously revisits the per-slice $\Delta_{s,m}$ values based on the set of active slices and network conditions to reposition each slice within the monitoring accuracy–overhead trade-off space. To show the benefit of such per-slice closed-loop control, this section compares it against (1) a \dint control plane (same threshold for all slices, similar to prior work \cite{sheng2021deltaint, chowdhury2021lint}), and (2) a \dintst control plane (different but fixed thresholds for each main slice type that does not change over time).
In our experiments, we set the monitoring error tolerance set to 5\% of the metric's \gls{SLA} value. Our goal is to show which control-plane strategy better navigates the trade-off between tolerance violations (i.e., accuracy) and bandwidth overhead.

\begin{figure*}[!ht]
\setlength{\belowcaptionskip}{-20pt}
    \centering
    \includegraphics[width=0.94\textwidth]{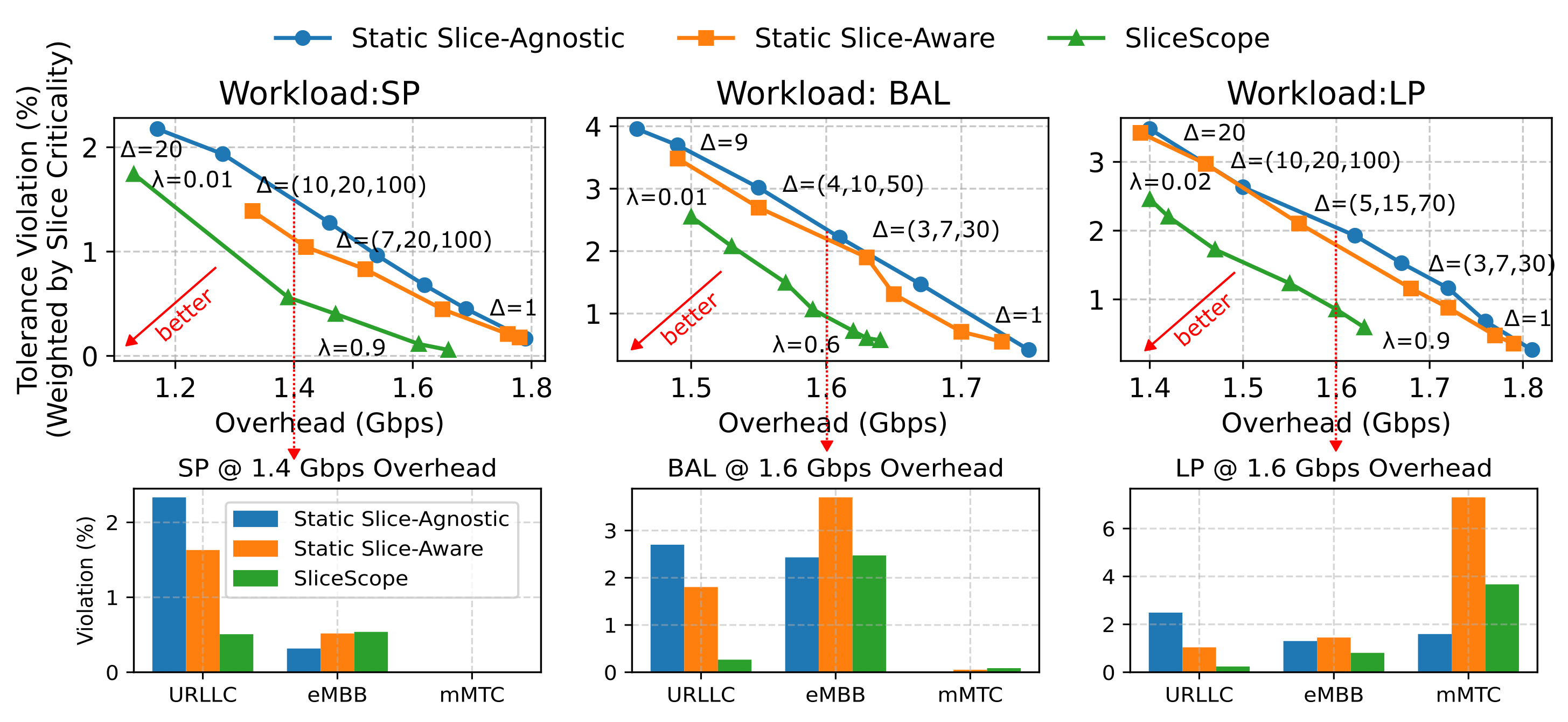}
    \caption{Pareto frontiers comparing per-packet overhead and tolerance violations across different workload mixes (SP, BAL, and LP). 
    \slicescope{} ``moves'' monitoring resources from less critical slices to more critical ones as needed.
    For Static Slice-Agnostic, points represent varying $\Delta$ values (1-20), for Static Slice-Aware, different combinations of $\Delta$ values for (URLLC, eMBB, mMTC) slices, and for \slicescope, different values of tuning parameter $\lambda$.}
    \label{fig:pareto-results}
\end{figure*}

Figure~\ref{fig:pareto-results} depicts the results. The Y axis represents the percentage of violations of the monitoring error tolerance across all packets, and the X axis is the total bandwidth overhead (lower and to the left is better).
For each control plane strategy, we plot the \emph{Pareto frontier} representing how it navigates this trade-off space. 
For \dint, each point is a different $\Delta$ value, which is fixed for all slices and over time. 
For \dintst, each point represents a combination of three specific $\Delta$ values, one for each of the three slice types. 
For \slicescope, each point represents a different value of the tuning parameter $\lambda$ in the optimization objective.

The results demonstrate that \slicescope{} achieves the best balance between overhead and tolerance violations across all workloads. While \dint with higher $\Delta$ values achieves lower overhead, it does so at the cost of increased tolerance violations, which can be particularly problematic for slices with tight \glspl{SLA}.
%
\dintst does better, showing the benefit of customizing $\Delta$ even just based on the main slice type, but fails to adjust to changing network conditions and active slices over time.
Specifically, for the BAL workload, \slicescope{} with $\lambda=0.6$ reduces tolerance violations by $\sim$2.3$\times$ compared to \dint{} and \dintst{}, while maintaining similar or slightly lower bandwidth overhead.

\indsignpost{A closer look per slice type} To better understand where \slicescope{} offers improvements, we break down the tolerance violations by slice type for each workload at a fixed bandwidth overhead (Fig.~\ref{fig:pareto-results}(bottom)). \slicescope{} consistently reduces violations for latency-sensitive URLLC slices compared to both \dint and \dintst across all workloads. For example, in the SP workload, \slicescope{} reduces the tolerance violations by 3$\times$ compared to the best performing static scheme. In the BAL workload, the reduction is up to 4.5$\times$.
This can come at the expense of less-critical slices. For example, In the LPS workload, \slicescope{} exhibits higher violation rates for mMTC slices. This behavior reflects its design: rather than uniformly allocating monitoring bandwidth to all traffic, \slicescope{} prioritizes critical slices where SLA violations are more costly, while reducing monitoring fidelity for less sensitive slices. In doing so, it allocates limited monitoring bandwidth resources where they matter the most.

\indsignpost{Take-away} The results underscore the importance of optimizing $\Delta$ values per slice. \slicescope{} effectively tracks critical slices with high accuracy, while achieving savings in overhead by assigning higher thresholds, i.e., fewer monitoring resources, to non-critical slices or in scenarios when the measured metric exhibits less variation. This adaptive approach enables \slicescope{} to balance the efficiency of selective INT insertion with the need for fine-grained monitoring of specific slices.

\subsection{Data Plane Primitives for Slice SLA Monitoring}
\label{sec:eval-e2e-tracking}

\begin{figure*}[!ht]
\setlength{\abovecaptionskip}{2pt}
\setlength{\belowcaptionskip}{-5pt}
    \centering
    \includegraphics[width=0.94\textwidth]{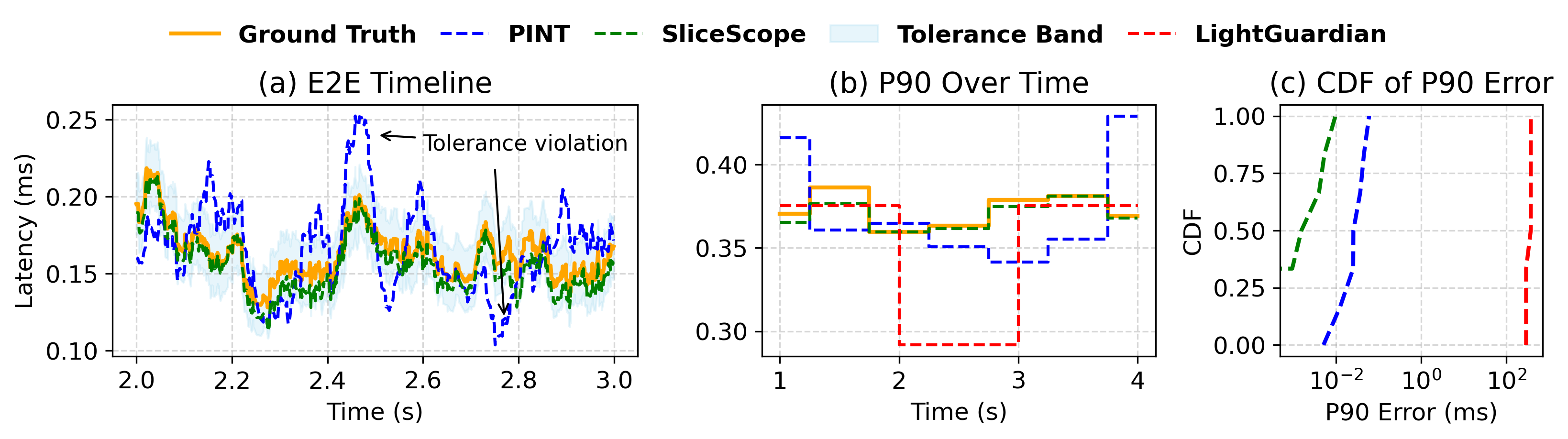}
    \caption{Comparison of different data-plane monitoring primitives for per-packet end-to-end slice monitoring}
    \label{fig:pint}
\end{figure*}

To validate our choice of threshold-based selective insertion for the data-plane of SLA-aware closed-loop control, we empirically compare \slicescope against two prominent 
alternative data plane primitives: \textit{probabilistic INT} (PINT 
\cite{pint2020probabilistic}) and \textit{sketch-based monitoring} (LightGuardian 
\cite{zhao2021lightguardian}). We evaluate their ability to provide accurate 
end-to-end tracking of per-slice SLA metrics, focusing on a scenario where an 
URLLC slice exhibits natural latency variation under our balanced (BAL) workload, as shown by the ground truth line in Figure~\ref{fig:pint}(a).
This comparison demonstrates why threshold-based insertion is superior for 
slice SLA monitoring compared to approaches that sacrifice per-packet 
end-to-end visibility. Similar trends are observed under the SP and LP workloads.

We compare the reported end-to-end latency from \slicescope{} and PINT at the per-packet level (Figure~\ref{fig:pint}(a)), and evaluate 90th percentile (P90) tracking accuracy across all three schemes -- \slicescope{}, PINT, and LightGuardian -- over 500ms intervals (Figure~\ref{fig:pint}(b)) and using the CDF of P90 error in ms (Figure~\ref{fig:pint}(c)). LightGuardian does not provide per-packet tracking and is therefore omitted from the timeline in Figure~\ref{fig:pint}(a). All schemes operate under comparable or higher bandwidth overhead budgets than \slicescope{}.

We observe that \slicescope{} consistently provides accurate tracking of both per-packet and aggregate (P90) latency, closely matching the ground truth. In contrast, PINT and LightGuardian show larger deviations: PINT has spikes where it violates the allowable error tolerances, and LightGuardian exhibits noticeable P90 error in end-to-end latency.

\indsignpost{Comparison with PINT \cite{pint2020probabilistic}} PINT implements selective INT for latency by a distributed sampling process where each packet probabilistically carries latency from one or more hops randomly chosen along its path. This allows lightweight tracking of per-hop latency quantiles under a given per-packet overhead budget. However, unlike \slicescope, it does not natively support per-slice end-to-end latency monitoring (\S\ref{subsec:why-delta-int}). 
To enable such comparison, we adapted PINT with a slice-aware reconstruction process at the collector: for each slice, the collector maintains the most recently observed latency for every hop, and reconstructs end-to-end latency by summing these values across the slice's path.
%
%
We compare the reported end-to-end latency from \slicescope{} and PINT at the per-packet level (Figure~\ref{fig:pint}(a)). PINT only reports per-hop latency quantiles. Due to the lack of per-packet alignment in PINT’s reconstruction of end-to-end latency (i.e., it may combine hop measurements from different packets in the same slice), its reported end-to-end latencies deviate significantly from the ground truth, as indicated by the two spikes crossing the allowable monitoring error tolerance.


\indsignpost{Comparison with LightGuardian \cite{zhao2021lightguardian}} Sketch-based approaches such as LightGuardian are designed to track per-hop distributions using compact, approximate data structures. LightGuardian uses the SuMax sketch \textemdash{} a variant of the count-min sketch (CMS), combined with binning, to track the distribution of per-hop latencies. It segments the latency range into multiple bins and uses CMS to count the number of packets in each bin. The sketches are then split into “sketchlets” and embedded into packets, allowing the collector to gradually reconstruct per-hop latency distributions.

To enable a fair comparison in our setting, we simulate LightGuardian with idealized conditions: we assume accurate per-hop sketches are received at the collector every 500ms, allowing the collector to reconstruct per-hop latency distributions for each slice.
Next, to compute end-to-end P90 latency for a slice, the collector convolves the per-hop latency distributions along the slice path to obtain the combined distribution, then extracts the 90th percentile. This convolution is necessary because LightGuardian does not track end-to-end metrics.
We also increase the number of bins from 4 to 10, compared to LightGuardian's original setup, to improve accuracy. 

Despite these favorable assumptions, LightGuardian's reconstructed slice-level end-to-end latency still deviates significantly from the ground truth (Figure~\ref{fig:pint}(b)), due to both the coarseness of its update interval and the inherent inaccuracy of deriving end-to-end metrics from per-hop metrics.
In contrast, \slicescope leverages the fact that the number of active slices is much smaller than the number of active flows to use a data-plane primitive that stores exact telemetry information (as opposed to approximate statistics) in the data plane and report that when there is a significant enough change, and it can directly provide accurate per-packet end-to-end latency.

\indsignpost{Take away} These experiments highlight the limitations of primitives such as per-hop sketches or probabilistic sampling approaches that do not natively support per-packet, end-to-end monitoring for SLA-aware closed-loop control of slice monitoring resources. Whether due to unaligned per-hop sampling or sketch-based summaries, these approaches struggle to provide accurate visibility into slice-level performance metrics.
\slicescope{} provides a better alternative in navigating the accuracy-overhead trade-off space in the presence of network slices with well-defined \glspl{SLA} where real-time and accurate tracking of end-to-end metrics is required.

\subsection{Microbenchmarks}
\label{sec:eval-microbenchmarks}

\signpost{Control plane scalability (optimizer and heuristic)} We evaluate the runtime of our control   
\begin{wrapfigure}{r}{0.52\textwidth}
    \centering
    \includegraphics[width=\linewidth]{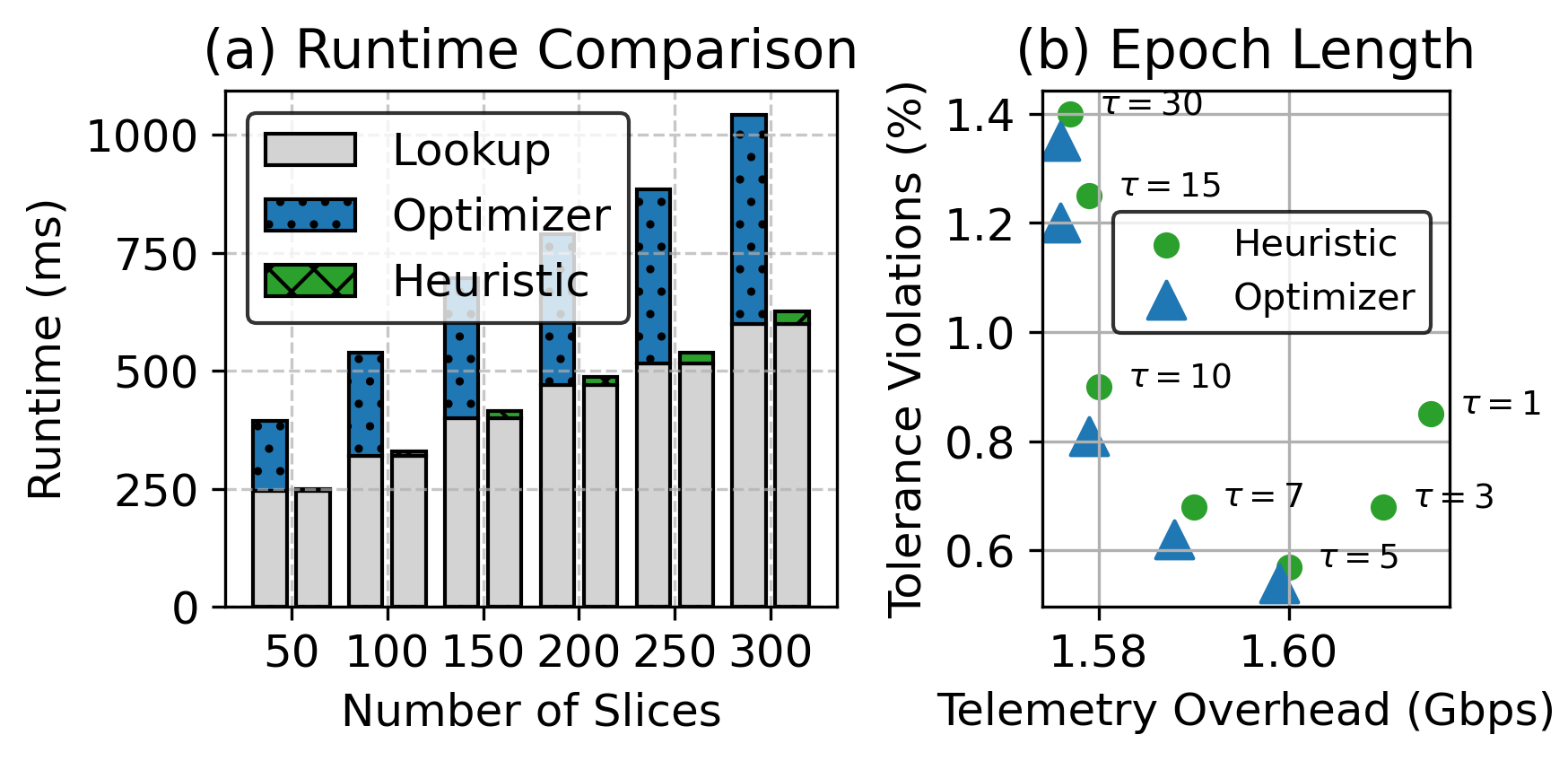}
    \caption{Control-plane scalability. (a) Runtime scaling with slice count for the ILP optimizer and heuristic fallback. (b) Monitoring quality across epoch lengths~$\tau$; the heuristic achieves comparable overhead–violation trade-offs while running $\sim$12× faster.}
    \label{fig:runtime_comparison}
    \vspace{-6pt}
\end{wrapfigure}
plane as the number of slices increases from 50 to 300. Recall from \S\ref{subsec:update-strategy} that \slicescope{} employs an ILP optimizer for threshold selection, with a greedy heuristic as a bounded-time fallback. Figure~\ref{fig:runtime_comparison}(a) shows how both strategies scale as the slice count increases. Each control epoch involves two stages: (1) computing expected monitoring overhead and error across candidate 
knobs $\Delta_{s,m}$ (\S\ref{subsec:math_framework}), and (2) selecting the final configuration via the optimizer or heuristic (\S\ref{subsec:optimization}).

To efficiently evaluate overhead and error across many candidate thresholds $\dknob$ (\S\ref{subsec:optimization}), \slicescope{} pre-computes and caches them in lookup tables.
This avoids repeatedly estimating $\beta(\dknob)$ and $E(\dknob)$ inside the optimizer: each candidate $\dknob$ would otherwise require a full recomputation of reporting probability and expected error, which the ILP solver (and even our heuristic) would invoke many times per iteration.
This lookup phase accounts for a significant portion of the total runtime and grows linearly with slice count, taking $\sim$250 ms at 100 slices and $\sim$500 ms at 300 slices on a 16-core machine. Although lookup cost scales linearly, it is inherently parallelizable since lookups for slice–metric pairs are independent and easily distributed across CPU cores.

Once lookup results are available, we select thresholds using either the ILP optimizer with early stopping or our greedy heuristic (\S\ref{subsec:update-strategy}).
The ILP solver's timeout is set proportional to the epoch length~$\tau$ to ensure the control plane adapts to network changes in a bounded time. For $\tau$ > $5$s, the optimizer consistently converges to high-quality solutions, while shorter epochs ($\tau < 5$s) more frequently trigger fallback to the heuristic.

As shown in Fig.~\ref{fig:runtime_comparison}(a), the optimizer adds $\sim$600 ms at 300 slices, while the heuristic incurs only $\sim$50 ms of additional latency.
\textit{Despite its $\sim$12× lower runtime, the heuristic achieves comparable overhead–violation trade-offs to the ILP} in regimes where both complete ($\tau \ge 5$s) (Fig.~\ref{fig:runtime_comparison}(b)).
This is because enforcing all monitoring-error constraints and minimizing overhead closely approximates the ILP's joint objective when constraints are tight, making the heuristic a practical, SLA-safe fallback under tighter time budgets.

\indsignpost{Epoch length} We also evaluate the effect of epoch length $\tau$ on bandwidth overhead and SLA tolerance violations (Fig.~\ref{fig:runtime_comparison}(b)). We observe that $\tau = 5$s achieves the best overall trade-off, with $\tau = 7$s a close second. Shorter intervals (e.g., 1–3s) provide insufficient time to learn stable metric variation patterns, leading to noisier $\Delta_{s,m}$ selection and higher overhead and violation rates. Longer intervals (e.g., 10–15s) reduce overhead by choosing more informed thresholds, but react too slowly to network changes, causing more frequent SLA violations. A 5–7s control interval offers a sweet spot: it is short enough to adapt to network dynamics, yet long enough to support accurate learning.

\begin{wrapfigure}{l}{0.50\textwidth}
    \centering
    \includegraphics[width=\linewidth]{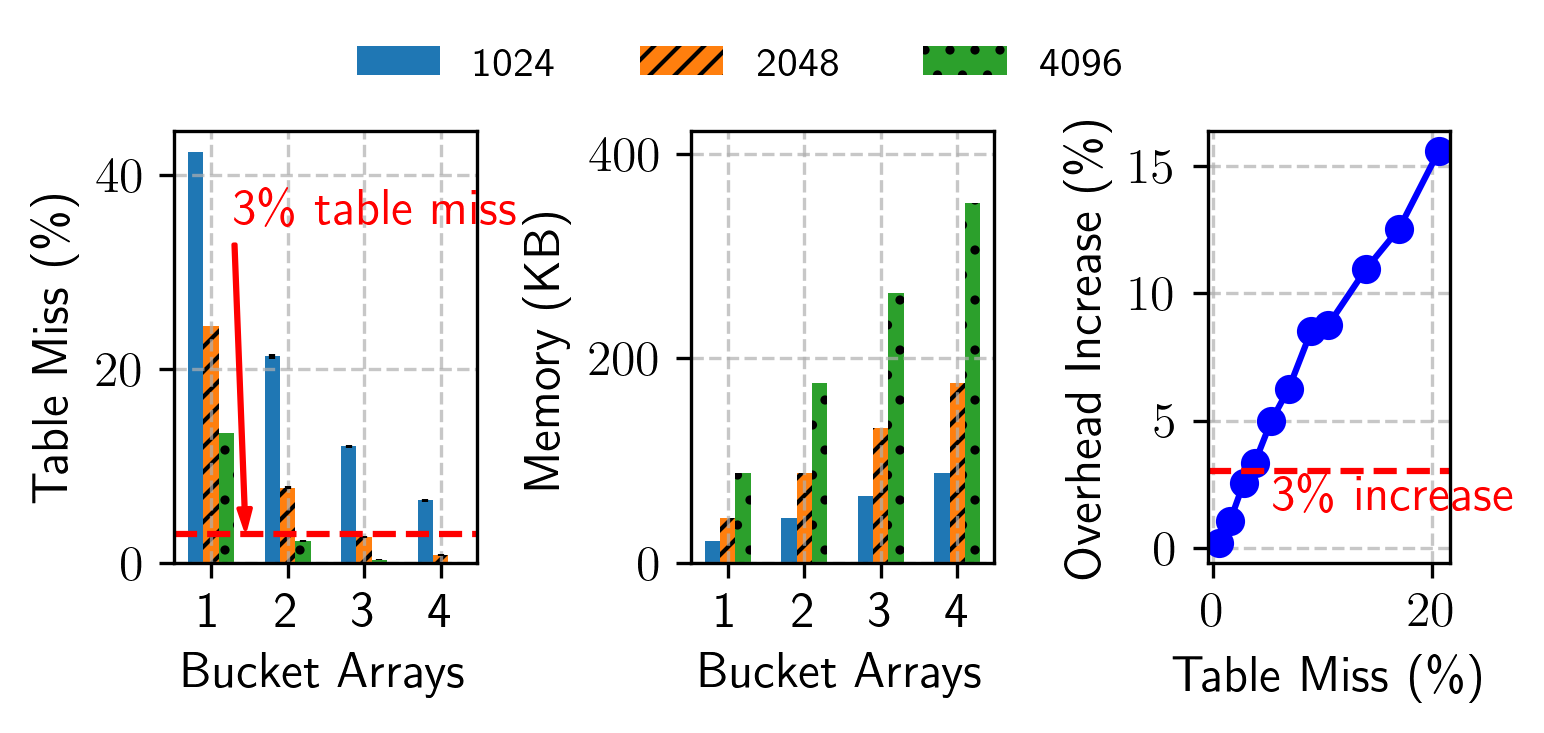}
    \caption{Bucket array sizing. Table-miss rate, memory usage, and overhead for different $(d,w)$ configurations. 
    A configuration of $d{=}2$, $w{=}4096$ achieves $<3\%$ miss rate with minimal extra bandwidth overhead.}
    \label{fig:bucket-array-sizing}
    \vspace{-6pt}
\end{wrapfigure}

\signpost{Bucket array sizing (Fig.~\ref{fig:bucket-array-sizing}}) \slicescope uses bucket arrays in the data plane to track per-slice metrics. When a lookup fails to find a match, i.e., a table miss, the upstream switch always emits the full telemetry in the next packet. This fallback ensures correctness but can increase overhead (\S\ref{subsec:corner_cases}). We vary the number of arrays ($d$) from $1$ to $4$ and number of buckets ($w$) from $1024$ to $2048$, and report, in Figure~\ref{fig:bucket-array-sizing}, 
the percentage of table miss (left), memory usage (middle), and associated increase in bandwidth overhead (right) with different configurations of $d$ and $w$. 
We observe diminishing returns as $d$ increases: from $d=1$ to $d=2$ table misses reduce significantly, but further increases yield only marginal gains while incurring higher memory cost. A configuration of $d=2$, $w=4096$ strikes a good balance, achieving <3\% table miss rate and only $\sim$3\% additional bandwidth overhead. Fewer arrays are preferable, as increasing $d$ consumes more data plane resources (Table~\ref{table:tofino-resource}).

\indsignpost{Summary} These microbenchmarks inform our system parameter choices: we set $\tau=5s$ for optimal overhead-violation trade-off and use the heuristic for scalability, and configure bucket arrays in the data plane with 4096 buckets and 2 arrays to meet our $<3\%$ miss rate target.

\subsection{Testbed Evaluation}
\label{sec:eval-testbed}

We evaluate our P4 implementation of \slicescope{} on a 5G hardware testbed comprised of: radio access network based on OpenAirInterface~\cite{openairinterface} and software-defined radios (SDRs), 5G core based on Open5GS~\cite{open5gs} and Intel Tofino switch (UfiSpace S9180-32X) as transport. A Google Pixel~7 Pro smartphone serves as the UE, enabling us to replicate real-world 5G traffic patterns. To further ensure realism, we replay commercial 5G traffic traces~\cite{korean-dataset} that include applications such as cloud gaming and live streaming. User-plane traffic is carried over GTP-U encapsulated flows, representative of slice traffic. To emulate multi-hop topologies within hardware limits, we apply external loopbacks on the Tofino, following prior work \cite{pint2020probabilistic}. 

\begin{figure}[!t]
\setlength{\belowcaptionskip}{-15pt}
    \centering
    \begin{minipage}[b]{0.58\textwidth}
    \setlength{\abovecaptionskip}{-5pt}
        \centering
        \includegraphics[width=\linewidth]{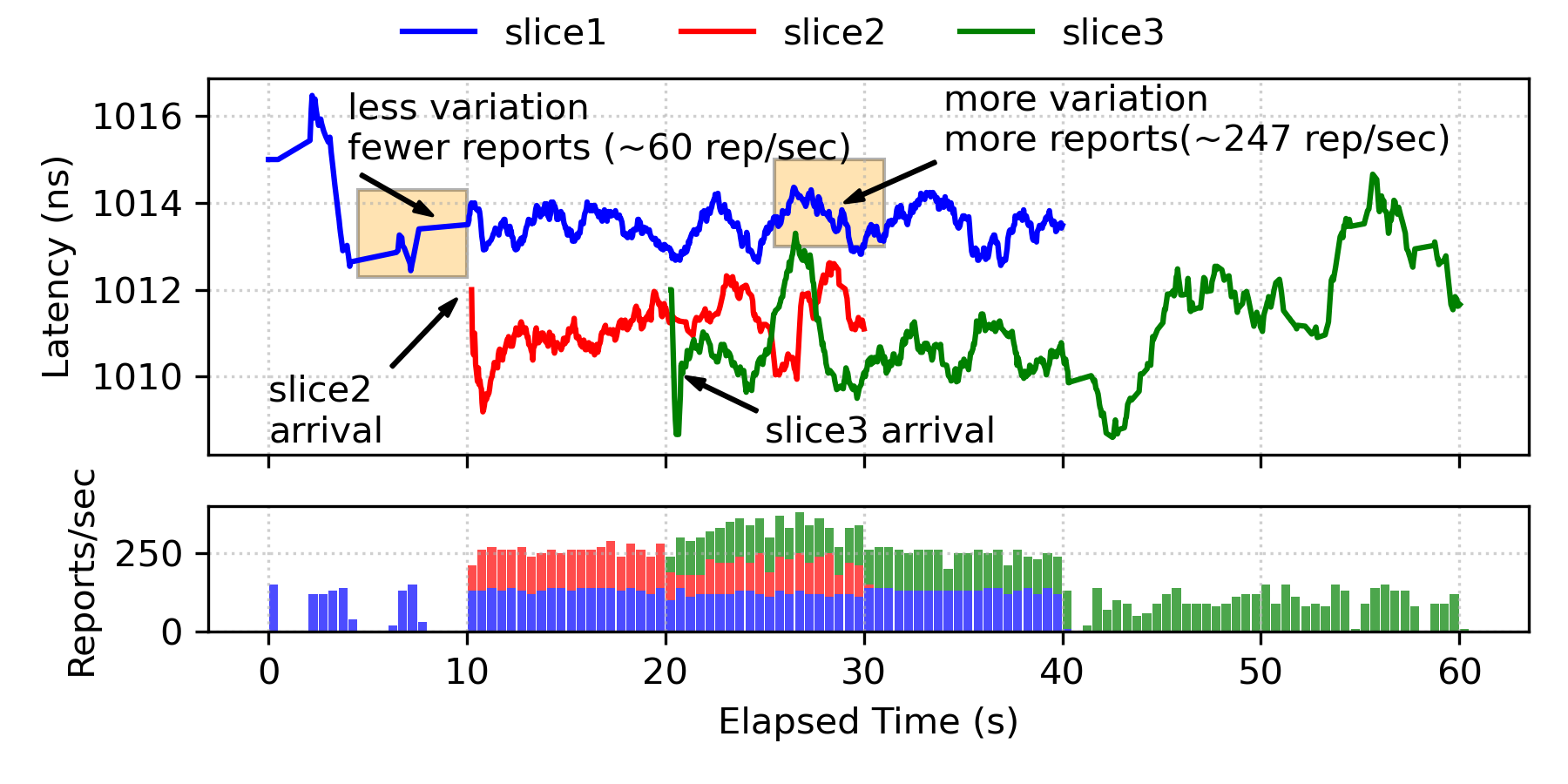}
        \caption{Tracking of slice metrics on Tofino.}
        \label{fig:tofino_timeline_pps}
    \end{minipage}
    \hfill
    \begin{minipage}[b]{0.40\textwidth}
    \setlength{\abovecaptionskip}{0pt}
        \centering
        \small
        \begin{tabular}{p{1.5cm}ccc}
            \toprule
            \textbf{Bucket Array $(d, w)$} & \textbf{Hash} & \textbf{SALU} & \textbf{SRAM} \\
            \midrule
            (1, 2048) & 7.2\% & 27.1\% & 9.2\% \\
            (1, 4096) & 8.4\% & 27.1\% & 9.2\% \\
            (2, 2048) & 11.1\% & 27.1\% & 9.4\% \\
            (2, 4096) & 12.5\% & 27.1\% & 9.4\% \\
            \bottomrule
        \end{tabular}
        \caption{Tofino resource usage with varying configurations. 
        For the selected setting $(d=2, w=4096)$, \slicescope{} consumes 12.5\% of hash units, 27.1\% of SALUs, and 9.4\% of SRAM.}
        \label{table:tofino-resource}
    \end{minipage}
\end{figure}

\indsignpost{Real-time tracking} Figure~\ref{fig:tofino_timeline_pps} presents real-time monitoring of slice E2E latency (top) and telemetry reporting rate (bottom) across three distinct slices. Traffic from slice 2 and 3 are started at second 10 and 20, respectively. On slice arrival, \slicescope immediately starts reporting metrics for the new slice, demonstrating real-time per-slice tracking.
The figure also shows that \slicescope adapts its reporting frequency in response to variation in the observed metric. While only slice 1 is active and latency is stable, the system reports at a low rate ($\sim$60 reports/sec), conserving bandwidth. After slice 2 and 3 join the network, latency variations increase, and \slicescope{} correspondingly raises its telemetry rate to $\sim$247 reports/sec. This behavior exemplifies the core design goal of \slicescope{}: providing real-time, selective telemetry that adjust telemetry bandwidth overhead according to SLA-relevant dynamics.

\indsignpost{Switch hardware overhead} Our prototype leverages three key hardware resources on the Tofino switch. 
\textit{Hash units} compute bucket indices for each row of the hash table, 
\textit{SRAM} stores the per-bucket telemetry state, 
and \textit{stateful ALUs (SALUs)} support one read/write per register array, with each row in the bucket array requiring one SALU. 
\slicescope{} occupies 11 pipeline stages, introducing only $\approx110$\,ns of additional processing delay per switch compared to simple forwarding. 
Resource usage for different bucket array sizes $(d,w)$ is summarized in Table~\ref{table:tofino-resource}.

\section{Conclusions and Future Work}

Modern networks increasingly rely on network slices with stringent real-time SLA requirements. Continuous monitoring is therefore essential, yet monitoring resources are limited. This work demonstrates that effective SLA-aware slice monitoring requires both adaptive control-plane mechanisms and suitable data-plane primitives. We formalized SLA-aware slice monitoring as a closed-loop control problem and introduced the Telemetry Primitive Contract (TPC) -- the minimal set of requirements a data-plane primitive must satisfy to enable such control. We then present \slicescope{}, a practical realization of this framework using change-triggered INT combined with adaptive, slice-aware control, which achieves up to $4\times$ better SLA tracking for critical slices compared to existing approaches.

More broadly, this work suggests viewing telemetry for network slices as a dynamically allocatable resource, managed through coordinated data and control plane mechanisms. This perspective opens new directions for adaptive, SLA-aware slice monitoring in programmable networks.

\emph{Future work.} \slicescope{} lays the groundwork for future SLA-driven telemetry systems. We are exploring more advanced control strategies, including learning-based and predictive mechanisms. We also plan to integrate \slicescope{} with real-time SLA enforcement and explore incorporating end-points that could be part of a slice (e.g., VNFs) using SmartNICs and/or eBPF.

\bibliographystyle{ACM-Reference-Format}
\bibliography{references}

\clearpage

\end{document}